\documentclass[onecolumn,traditabstract]{aa}
\usepackage{graphicx}
\usepackage{natbib}
\usepackage{epsfig}
\usepackage{lscape}
\usepackage{graphicx}
\usepackage{amsmath}
\usepackage[english]{babel}
\usepackage[titletoc,title]{appendix}
\usepackage{txfonts}
\usepackage{times}

\begin{document}
\title{Triggering Active Galactic Nuclei in Hierarchical Galaxy Formation: Disk instability vs. Interactions}

  \author{N. Menci, M. Gatti, F. Fiore, A. Lamastra}
   \institute{INAF - Osservatorio Astronomico di Roma, via di Frascati 33, 00040 Monte Porzio Catone, Italy.}
   \abstract{Using a state-of-the-art semi analytic model for galaxy formation we have investigated in detail the effects of Black Hole (BH) accretion triggered by disk instabilities (DI) in isolated galaxies on the evolution of the AGN population. Specifically, we took on, developed and expanded the Hopkins \& Quataert (2011) model for the mass inflow following disk perturbations, based on a physical description of nuclear inflows and  tested against aimed N-body simulations. We have compared the evolution of AGN due to such a DI accretion mode with that arising in a scenario where galaxy interactions (IT mode) produce the sudden destabilization of large quantities of gas feeding the AGN;  this constitutes the standard AGN feeding mode implemented in the earliest versions of most semi-analytic models. To study  the {\it maximal} contribution of DI to the evolution of the AGN population, we extended and developed the Hopkins \& Quataert (2011) model to assess the effects of changing the assumed disk surface density profile, and to obtain lower limits for the nuclear star formation rates associated to the DI accretion mode. We obtained the following results: i) for luminosity $M_{1450}\gtrsim -26$ AGN the DI mode can provide the BH accretion needed to match the observed AGN luminosity functions up to $z\approx 4.5$; in such a luminosity range and redshift, it constitutes a viable candidate mechanism to fuel AGN, and can compete with the IT scenario as the main driver of cosmological evolution of the AGN population: ii) The DI scenario cannot provide the observed abundance of high-luminosity QSO with $M_{1450}\leq -26$ AGN, as well as the abundance of 
high-redhshift $z\gtrsim 4.5$ QSO with $M_{1450}\leq -24$.  As found in our earliest works, the IT scenario provides an acceptable match to the observed luminosity functions up to $z\approx 6$; iii) The 
 dispersion of the distributions of Eddington ratio $\lambda$ for low- and intermediate-luminosity AGN (bolometric $L_{AGN}=10^{43}-10^{45}$ erg s$^{-1}$) is predicted to be much smaller in the DI scenario compared to the IT mode; iv) The above conclusions concerning the DI mode are robust with respect to the explored variants of the Hopkins \& Quataert (2011)  model. We discuss the physical origin of our findings.  Finally, we discuss how it is possible to pin down the dominant fueling mechanism of AGN 
in the  low-intermediate luminosity range $M_{1450}\gtrsim -26$ where the DI and the IT modes are both viable candidates as the main drivers 
of the AGN evolution. We show that an interesting discriminant could be provided by the fraction of AGN with high Eddington ratio $\lambda\geq 0.5$, since it  increases with luminosity in the IT case, while the opposite is true in the DI scenario. 

}
\authorrunning{N. Menci, et al.}
 \maketitle

\section{Introduction}

Understanding the processes driving the growth and evolution of the Supermassive Black Holes (BHs) constitutes a major goal in astrophysics. At the beginning of the 
last decade, it become widely accepted that the such an evolution is strongly correlated with that of the host galaxies, due to the observed tightness of the relations between the BH mass and the galaxy properties such as the stellar mass or the velocity dispersion of the bulge (Kormendy \& Richstone 1995; Magorrian et al. 1998, Ferrarese \& Merritt 2000; Gebhardt et al. 2000; Marconi \& Hunt 2003; H\"aring and Rix 2004; Aller \& Richstone 2007). Together with the observed low fraction of Active Galactic Nuclei 
(AGN) and with the strong cosmological evolution of the number density of bright AGNs (see, e.g., Richards et al. 2006; Croom et al. 2009; Ueda et al. 2014; for a review see Merloni \& Heinz 2013), such measured relations strongly indicated that a major fraction of the BH mass results from accretion of gas during relatively short (duration $\tau\approx 10^7$ yrs) but intense (accretion rates $\dot M_c\approx 1-10$ $M_{\odot}$/yr), repeated  bursts (see, e.g., Shankar, Weimberg, Miralda-Escude' 2009; Fiore et al. 2012). Given such a framework, the focus has progressively shifted toward understanding the triggering mechanisms responsible for such intense and short accretion episodes in the galaxy nuclei, and on their relation with the cosmological evolution of galaxies. However, connecting the nuclear accretion with the larger-scale properties of galaxies is not an easy task: even the matter located at a radius of a few hundreds of pc from the central BH must lose more than 99.99\% of its angular momentum to be effectively accreted. Numerical simulations on galactic scales have shown that gravitational torques induced by mergers (Hernquist 1989, Barnes \& Hernquist 1991; 1996; Heller \& Shlosman 1994 ) or  instabilities in self-gravitating disks (Hernquist \& Mihos 1995; Bournaud et al. 2005; Eliche-Moral et al. 2008; Younger et al. 2008) - especially in gas rich, high-redshift galaxies (Bournaud et al. 2007; Genzel et al. 2008) - are effective in causing gas inflow down to scales $\lesssim 1$ kpc. However, both theoretical arguments (Schwarz 1984; Combes \& Gerin 1985; Shlosman et al. 1989; Athanassoula 1992) and observational results (see Jogee 2006 for a review) strongly indicate that the transport of gas from  galactic scales down to the pc scale (where the BH gravity dominates) requires additional physics. Again, both external and internal triggers may be effective in providing additional loss of angular momentum. In the case of major mergers, as violent relaxation starts, the gas experiences rapidly varying gravitational torques and suffers strong shocks which can effectively result into large gas inflows in the central region (see Hopkins et al. 2010 and references therein). In the case of isolated galaxies, recent simulations show that - as the disky material at smaller radii becomes self-gravitating - the large-scale bars might indeed trigger a cascade of secondary gravitational instabilities that transport the gas toward the centre (Hopkins \& Quataert 2010); the latter process resembles the "bars within bar" scenario (see Shlosman et al. 1989; Friedli \& Martinet 1993; Heller \& Shlosman 1994; Maciejwski \& Sparke 2000) but includes a wider range of morphologies in the nuclear region. 

Assessing the role of such processes in determining the cosmological co-evolution of AGN and galaxies is a challenging task.  

On the observational side, the situation is complex. 
On the one hand, a correlation between strong galaxy interactions and the AGN activity is found in very luminous QSOs 
with high mass accretion rates $\gtrsim 10$ M$_{\odot}$ yr$^{-1}$ (Disney et al. 1995; Bahcall et al. 1997; Kirhakos et al. 1999; Hutchings 1987; Yates et al. 1989; Villar-Martín 2010; 2012; Treister et al. 2012); recently, a clear statistical increase in the AGN fraction in close pairs of galaxies has been found by Ellison et al. (2011), while   minor mergers as triggers for lower luminosity AGNs has been proposed to explain the observational features in some 
nearby galaxies (see, e.g., Combes et al. 2009); the discovery of double AGNs (e.g. Hennawi et al. 2010; McGurk et al. 2011) adds support to a merger origin of the AGN activity. 
On the other hand, 
various observational studies suggest that moderately luminous AGN are typically not major-merger driven at $z \lesssim 1$ (Salucci et al. 1999; Grogin et al. 2005; Pierce et al. 2007; Georgakakis et al. 2009;  Cisternas et al. 2011; Villforth et al. 2014), and interestingly also at $z \approx 2$ (Rosario et al. 2013a; Silverman et al. 2011; Kocevski et al. 2012); Bournaud et al. (2012) found a correlation between the presence of violent instabilities and AGN activity in gas rich galaxies at high redshift, due to the presence of giant clumps of gas in the disk.

On the theoretical side,  aimed, high-resolution N-body simulations can study specific galaxy systems, but understanding the relative effects of the disk instability (DI hereafter) and of the galaxy interactions (IT thereafter) as AGN feeding modes requires implementing the physics of nuclear gas inflows into cosmological models of galaxy formation. In turn, this requires an analytical description of such processes to be implemented into existing semi-analytic models (see Baugh 2006 for an introduction), or as sub-grid physics in cosmological simulations. As for the IT scenario, multiscale high-resolution hydrodynamical N-body simulations are now resolving scales $\sim 10$ pc, showing that collisions efficiently cause most of the gas to flow to the centre on timescales of a few dynamical times (see Hopkins et al. 2010 and references therein); analytical- though simplified - or phenomenological descriptions of the inflow determined by mergers have been developed by several authors (see, e.g., Makino \& Hut 1997, Cavaliere \& Vittorini 2000), calibrated and  tested against binary hydrodynamic merger simulations (Robertson et al. 2006a,b; Cox et al. 2006; Hopkins et al. 2007b); here the fraction of galactic gas which is able to loose angular momentum and flow to the centre is related to large relative variations $\Delta j/j$ of the gas specific angular momentum induced by the interactions, which is in turn connected to the mass ratio $M'/M$ of the merging partners. A different situation holds  for the DI scenario. Until recently, no analytic theory that accurately predicts the inflow rates of gas in mixed stellar-plus-gas systems has been worked out; in fact, most of the existing works have been based on the over-simplified assumption of completely stellar or gaseous disks (see, e.g., Goldreich \& Lynden-Bell 1965; Lin et al. 1969; Toomre 1969;  Lynden-Bell and Kalnajs 1972; Goldreich \& Tremaine 1980; Binney \& Tremaine 1987; Shu et al. 1990; Ostriker et al. 1992). In fact, so far, the inflows driven by disk instability have been implemented into semi-analytic models only through {\it ansatz}  concerning the fate of disk gas and stars when the the disk becomes unstable; e.g., the model by Hirshman and Somerville (2012) assumes that a fixed (and tunable) fraction of the gas mass exceeding the threshold for instability is accreted onto the BH, while Fanidakis et al. (2012) assume that all gas and stars in the disk are transferred to the bulge component, leading to more dramatic effects. Due to the lack of an analytical description of the complex physics processes acting in the transport of gas in the nuclear region, the results strongly depend on the assumed prescriptions and may actually overestimate the real inflows attainable in the central regions under realistic conditions.

Recently, however, a step forward in the description of the gas inflow due to disk instabilities has been proposed by Hopkins \& Quataert (2011; HQ11 hereafter); the authors develop an analytic model that attempts to account for the physics of angular momentum transport and gas inflow from galactic scales ($\approx 1$ kpc) inward to the nuclear scales $\sim 10$ pc where the BH potential is dominant. The multi-component nature of the disk is considered, and the inflow toward the BH is related to the nuclear star formation; such  a model has tested by the authors against aimed hydrodynamical simulations, so that at present it constitutes a solid baseline to describe the BH accretion due to disk instabilities. 

In this work, we include the HQ11 model for disk instability in an updated version of the Rome semi-analytic model of  galaxy formation (Menci et al. 2006, 2008) to compute the 
statistical effects of such an accretion mode on the AGN population, within a cosmological context; the results are compared with those obtained assuming interactions as the only trigger of AGN activity (our reference model). It is important to emphasize that here we do not attempt to develop a best-fitting model, tuning the relative role 
of the two accretion triggers; in this work our aim is to study the {\it maximal} effects of a physical model for BH accretion triggered by disk instabilities onto the AGN and BH statistics, and to enlighten the main differences with respect to the predictions of purely interaction-driven models for AGN feeding. The final goal - after comparing with observations - is to single out the physical regimes where the different feeding modes may be effective. 

The paper is organized as follows: in Sect. 3 we recall the main properties of the semi-analytic model that we use, enlightening the modifications with respect to the previous version; we also test the updated model by comparing the predicted evolution of the galaxy population with existing observations. In Sect. 4 we describe how we implement different BH accretion mechanisms into our semi-analytic model; in Sect. 4.1 we recall our reference model based on galaxy interactions, while in Sect. 4.2 we present our implementation of the HQ11 model for accretion from disk instabilities. We also present extensions of the HQ11 model and on variants obtained by relaxing or modifying some of the assumptions of the original HQ11 model. We present our results in Sect. 5, while Sect. 6 is devoted to discussion and conclusions.

\section{The Semi-Analytic Model}

We base on the semi-analytic model described in Menci et al. (2006, 2008), to which we refer for details; here we recall its key points and we describe the main updates. 
The backbone is constituted by the merging trees of dark matter haloes, generated through a Monte Carlo procedure on the basis of the merging rates given by the Extended Press \& Schechter (EPS, see Bardeen, Bond, Kaiser, Szalay 1991; Lacey \& Cole 1993; Bower 1991) formalism. After each merging event, the dark matter haloes
 included into a larger object may survive as satellites, or sink to the centre due to dynamical friction to increase the mass of the central dominant galaxy. 
The baryonic processes taking place into each dark matter halo are then computed; the gas in the halo, initially set to have a density given by the universal baryon fraction and to be at the virial temperature, cools due to atomic processes and settles into a rotationally supported disk with mass $M_c$, disk radius $R_d$ and disk circular velocity $V_d$ computed as in Mo, Mao \& White (1998). The cooled gas $M_c$ is gradually converted into stars, with a rate $\dot M_*=M_c/\tau_*$ given by the Schmidt-Kennicut law with $\tau_*=1$ Gyr.  In addition to the above "quiescent" mode of star formation, galaxy interactions occurring in the same host dark matter halo may induce the sudden conversion of a fraction $f$ of cold gas into stars on a short time scale $\sim 10^7-10^8$ yrs given by the duration of the interaction. 
The fraction $f$ is related to the mass ratio and relative velocity of the merging partners as described in Menci et al. (2003).  
The energy released by the Supernovae associated to the total star formation returns a fraction of the disk gas into the hot phase,  providing the feedback needed to prevent over cooling. An additional source of feedback is provided by the energy radiated by the AGN which correspond to the active accretion phase of the central BH; 
the detailed description of our implementation of the AGN feedback is given in Menci et al. (2008). In our reference model, the triggering and the luminosity of AGN are 
determined by galaxy interactions as described in Sect. 4.1. Finally, the luminosity - in different bands - produced by the stellar populations of the galaxies are computed by convolving the star formation histories of the galaxy progenitors with a synthetic spectral energy distribution, which we take from Bruzual \& Charlot (2003) assuming a Salpeter IMF. 

With respect to the model version described in our previous works, we performed some changes that we describe below. First, we include a treatment of the transfer of stellar mass to the bulge during mergers. A canonical prescription in semi-analytic models (see Khochfar \& Silk 2006; Parry Eke \& Frenk 2009; Benson \& Devereux 2010; De Lucia et al. 2011) is to 
assume that the stellar content of galaxies is transferred to the bulge during major mergers characterized by  stellar mass ratios $\mu=M_{*1}/M_{*2}$ of the merging partners larger than some fixed fraction $0.2-0.3$ calibrated through N-body simulations. Recently, Hopkins et al. (2009) showed that the distribution of the bulge-to-total (B/T) ratio provides a better match to the observations when the transferred stellar mass in mergers is a function of both merger mass ratio and gas fraction in the progenitor galaxies. Following the latter authors, we assume that in mergers with $\mu\geq 0.2$ only a fraction $1-f_{gas}$ of the disk mass is transferred to the bulge. 
The resulting $B/T$ distribution is compared with observations from Fisher \& Drory (2011) and 
Lackner \& Gunn (2012) in fig. 1. Note, however, that our following results (Sect. 5) will not change appreciably if we take the canonical assumption that all disk stars are transferred into the bulge during major mergers. A second modification of our original model 
is the introduction of a description of the the tidal stripping of stellar material from satellite galaxies. We adopt exactly the treatment introduced by  Henriques \& Thomas (2010) in the Munich semi-analytic model, to which we refer for details. Here we just recall that the model is based on the computation of the tidal radius for each satellite galaxy; this is identified as the distance from the satellite centre at which the radial forces acting on it cancel out (King 1962; Binney \& Tremaine 1987; see also Taylor \& Babul 2001). These forces are the gravitational binding force of the satellite, the tidal force from the central halo and the centrifugal force. In the simple approximation of nearly circular orbits and of an isothermal halo density profile,  such a radius can be expressed as $r_t\approx \sigma_{sat}\,r_{sat}/\sqrt{2}\,\sigma_{halo}$, where $\sigma_{sat}$ and $\sigma_{halo}$ are the velocity dispersions of the satellite and of the halo, respectively, and $r_{sat}$ is the halocentric 
radius of the satellite, that we computed in our Monte Carlo code during its orbital decay to the centre due to dynamical friction. For each satellite galaxy, the material outside this radius is assumed to be disrupted and becomes a diffuse stellar component in the host halo (see Henriques \& Thomas 2010). 

The above model constitutes our  baseline to connect the growth of BHs to the properties of their host galaxies. Thus, we first test the updated version of our model to ensure it provides a reasonable account of the galactic properties and of their evolution over a wide range of cosmic times, assuming a concordance cosmology (see Spergel et al. 2006). In fig. 1 we show how the model predictions compare with different local observables: the stellar mass function, the B-band luminosity function, and the ratio of the bulge to the total stellar mass. The good matching indicates that model provides a fair description of both the (roughly) instantaneous and integrated star formation rates, and that our treatment of stellar mass stripping and morphological transformations is consistent with observations. 

\begin{center}
\vspace{-0.4cm}
\scalebox{0.30}[0.30]{\rotatebox{0}{\includegraphics{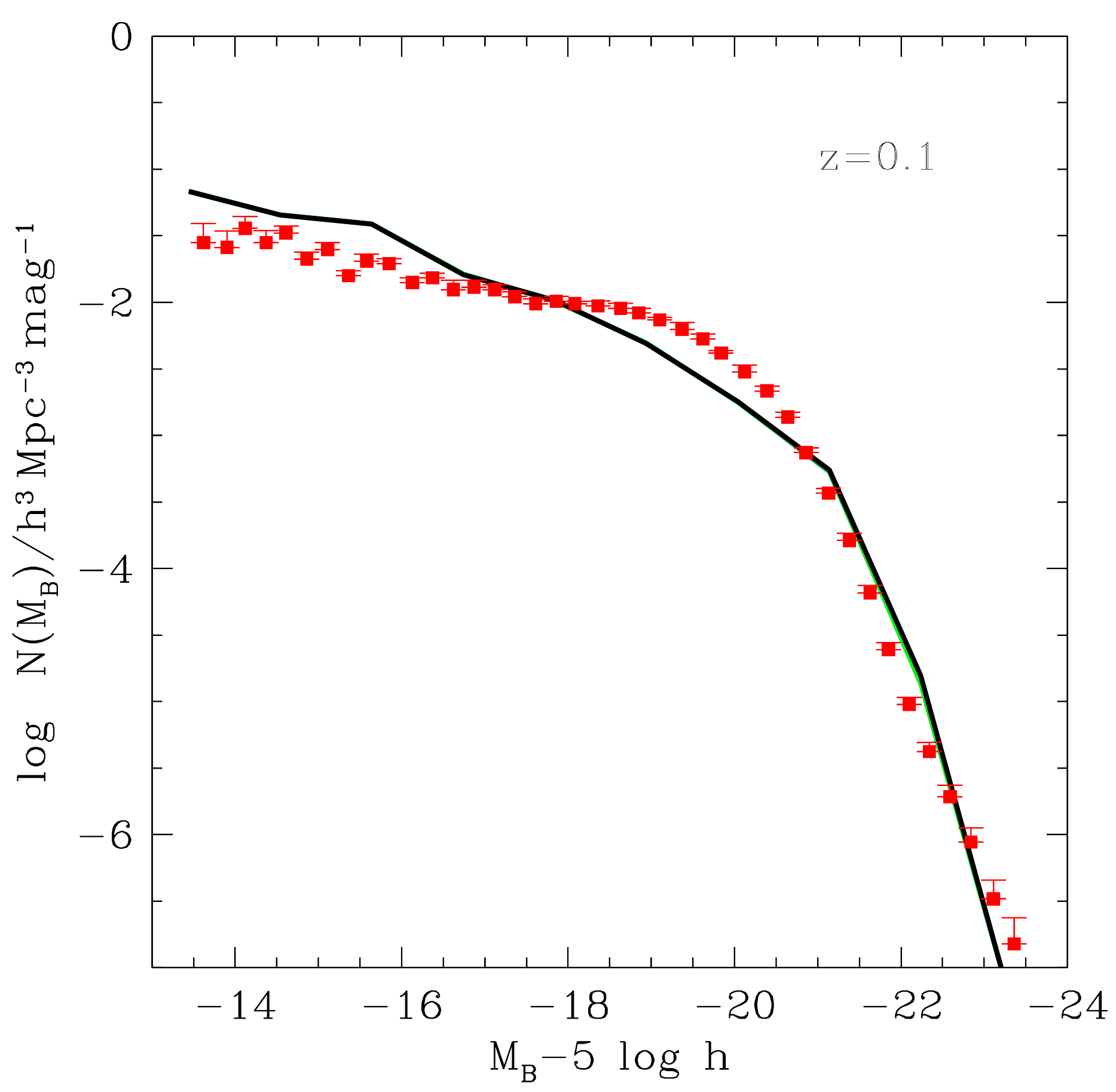}}}
\scalebox{0.30}[0.30]{\rotatebox{0}{\includegraphics{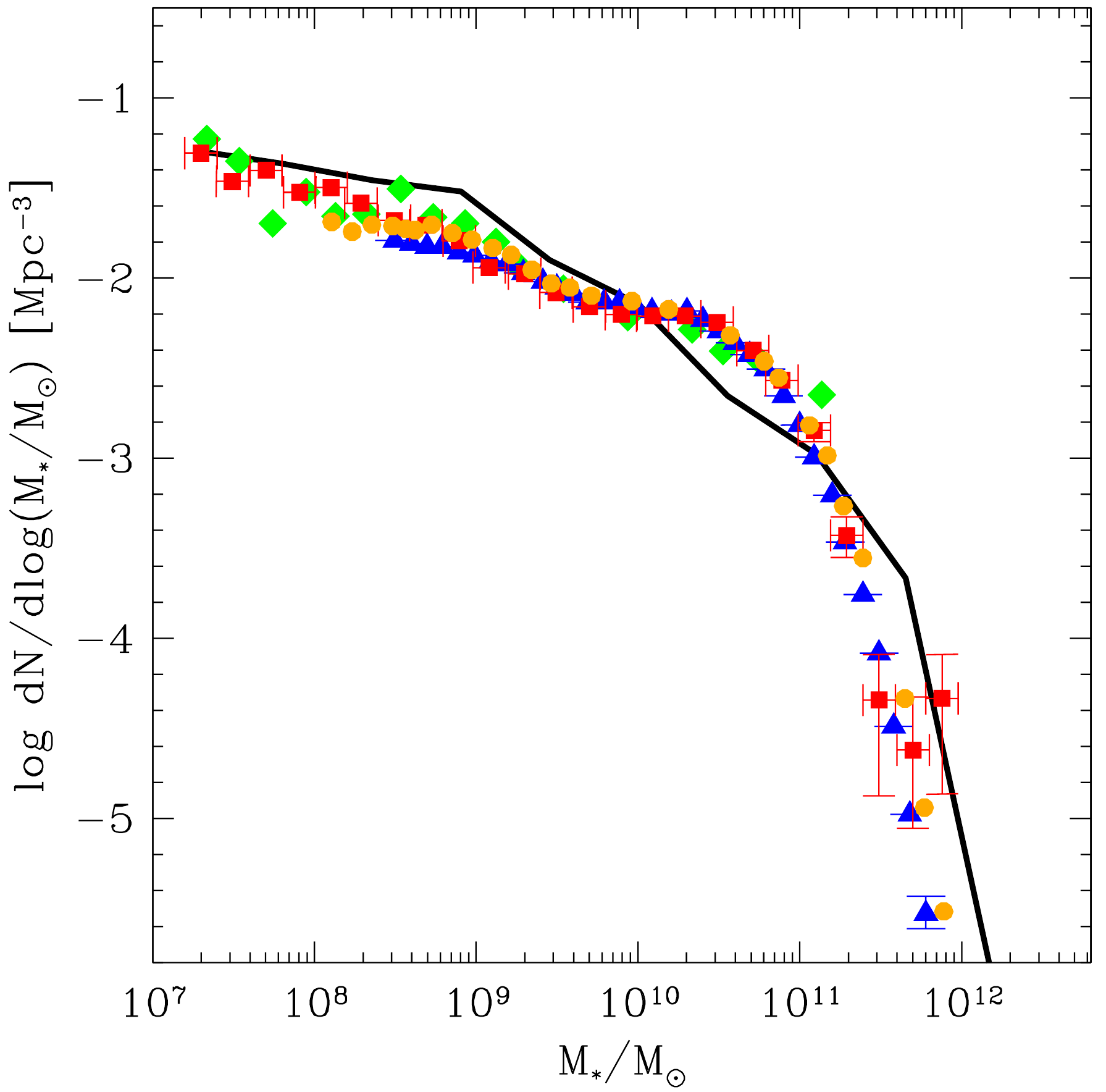}}}
\scalebox{0.36}[0.38]{\rotatebox{0}{\includegraphics{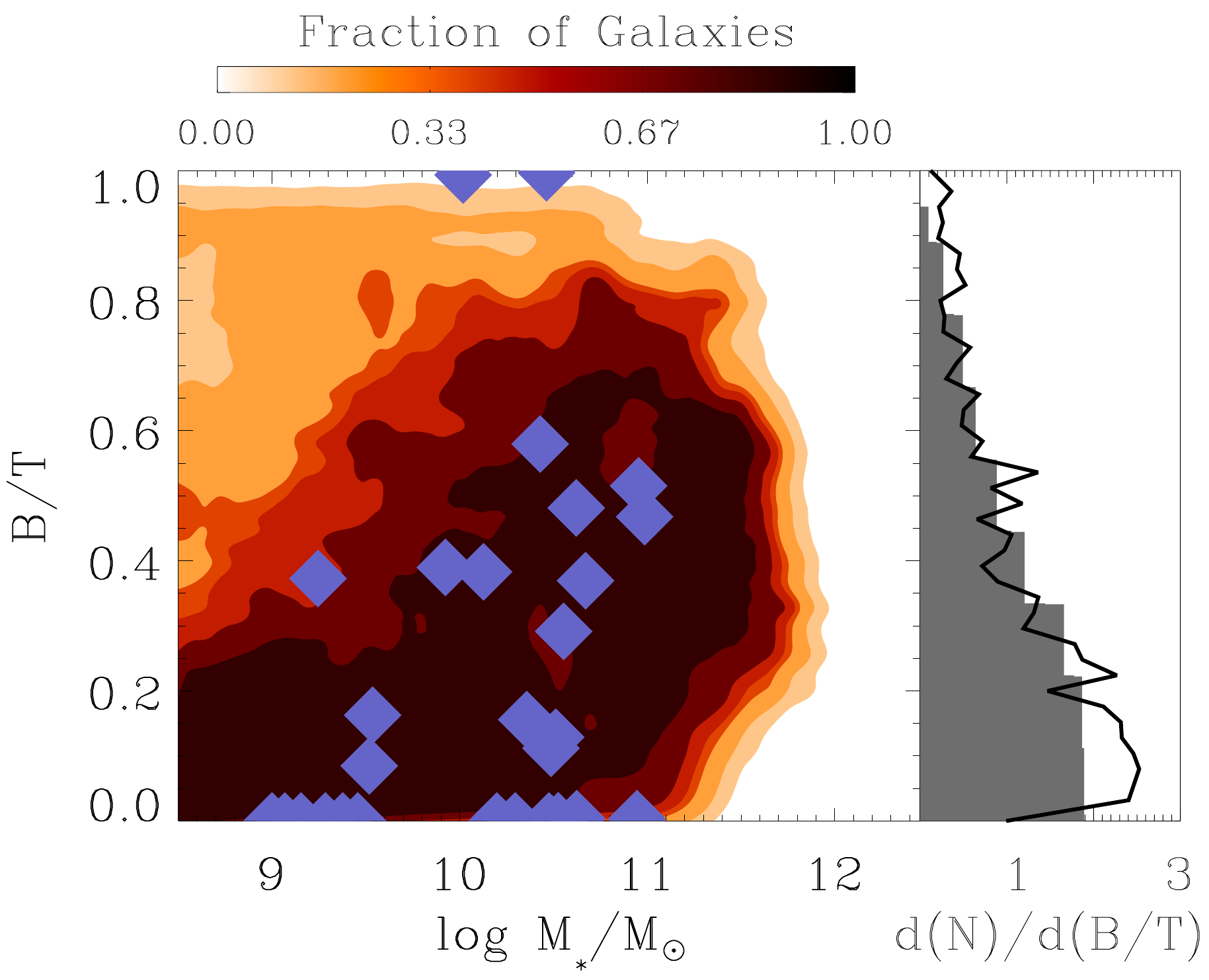}}}
\end{center}
\vspace{-0.4cm }
 {\footnotesize 
Fig. 1. - Left panel: The predicted luminosity function at $z\leq 0.1$ (in bj band, solid line) is compared with observations from Jones et al. (2006). 
Central Panel: The predicted local stellar mass function (solid line) is  compared with data from Baldry et al. (2012, squares), Li \& White (2009, triangles), Panter et al. (2007, circles),  Fontana et al. (2006, diamonds). 
 Right Panel: The contour plot shows the predicted relative fraction of bulge-to-total mass as a function of the galaxy stellar mass is compared with data from Fisher \& Drory (2011); the corresponding distribution of B/T ratios (differential number of objects per B/T bin, normalized to the total number, solid line) is also compared with the SDSS data (Lackner \& Gunn 2012, shaded histogram) in the rightmost panel, for galaxies in the 
magnitude range $-24\leq M_r\leq -17$.  
\vspace{0.4cm}}

As a further test we show in fig. 2 (left panel) 
 the predicted local color-magnitude distribution, which seems to capture the well known bimodal nature of the observed distribution. To quantify such a result, we compare in detail 
 the predicted color distribution of galaxies with different magnitudes (the right panel of fig. 2) 
 with the SDSS (Sloan Digital Sky Survey) data; besides the mentioned color bimodality, the fraction of red galaxies  continuously increases with increasing galaxy magnitude, as indicated by the data. 
 
\begin{center}
\vspace{-0.cm}
\scalebox{0.44}[0.44]{\rotatebox{-0}{\includegraphics{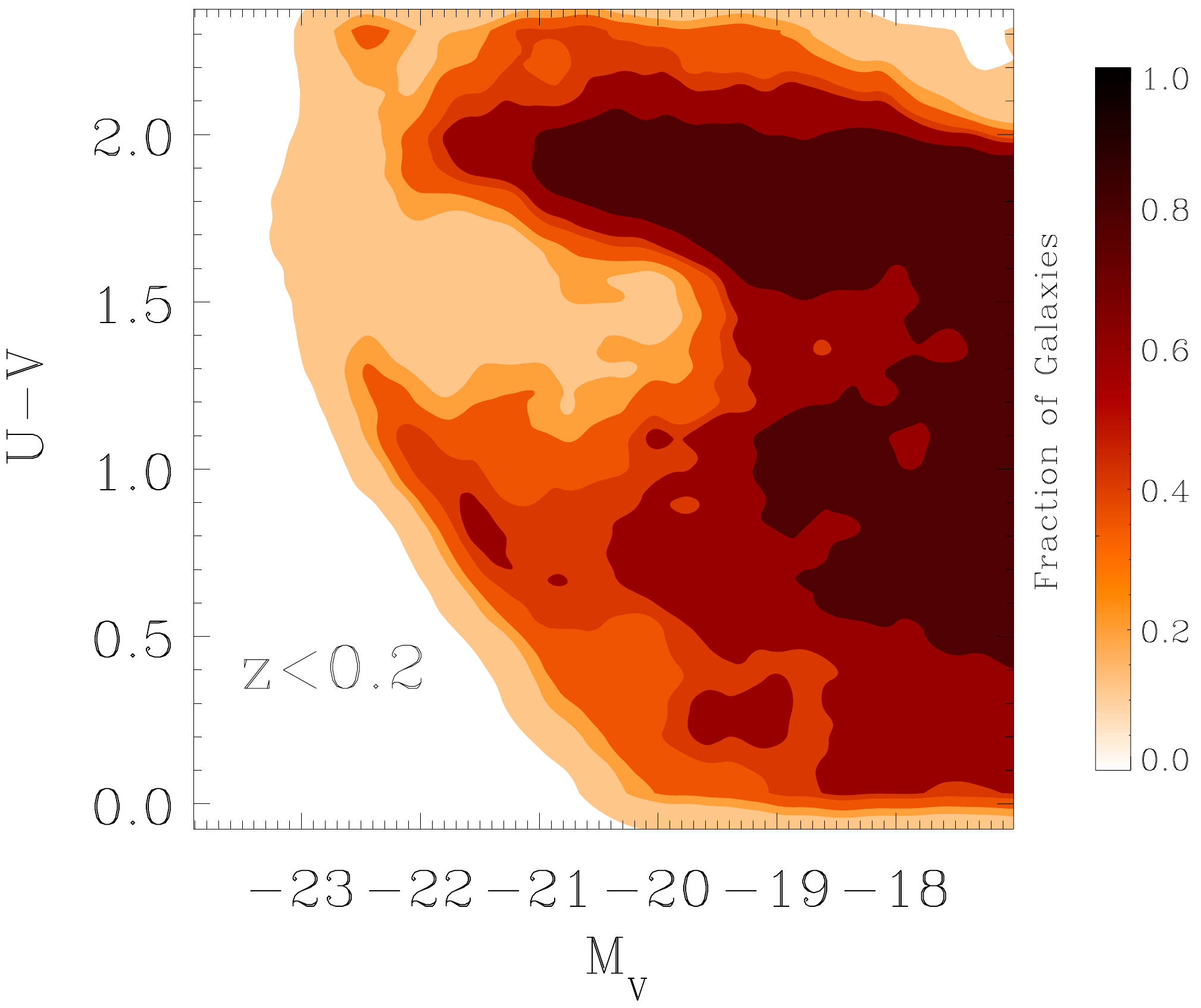}}}
\hspace{1cm}
\scalebox{0.3}[0.3]{\rotatebox{-0}{\includegraphics{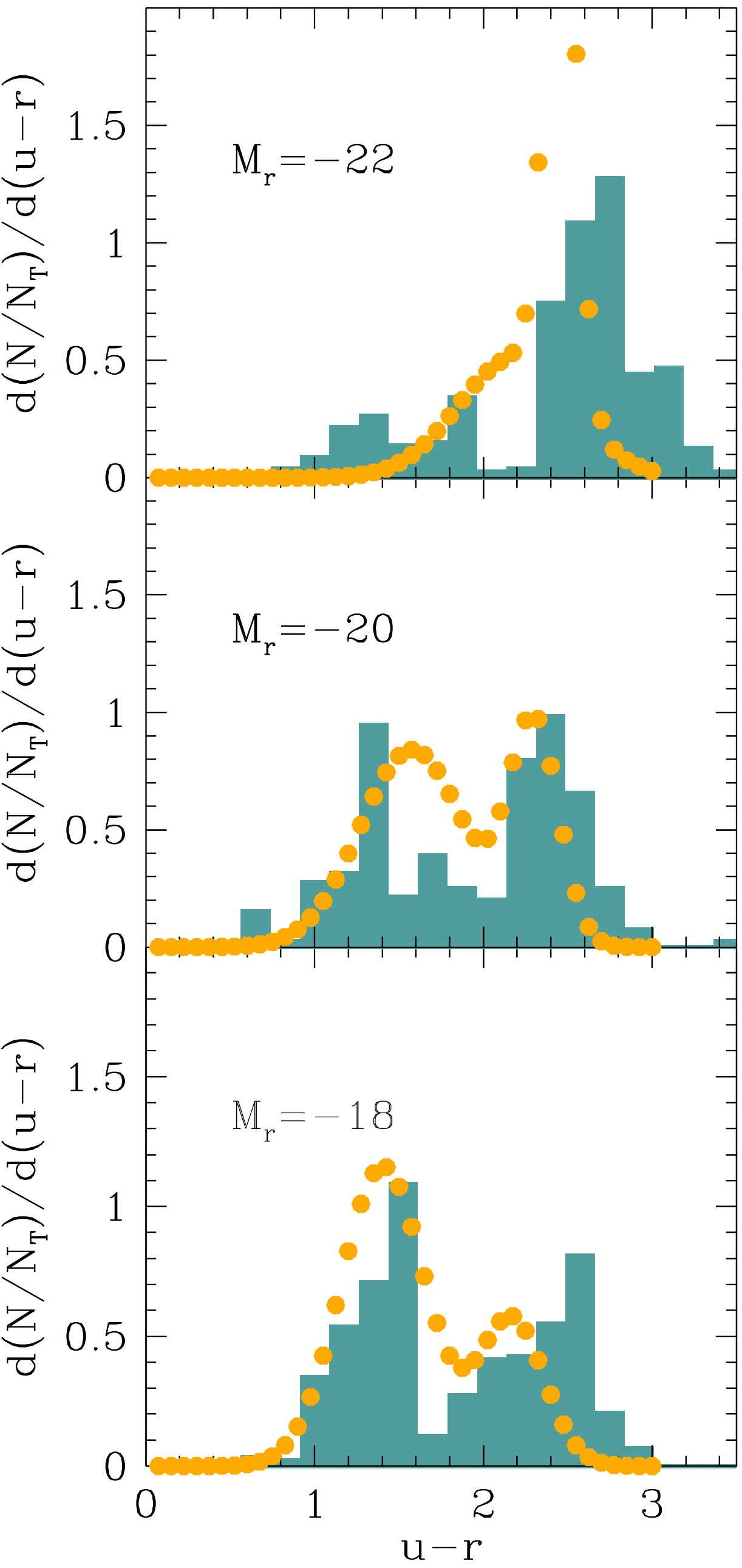}}}
\end{center}
\vspace{-0.2cm }
 {\footnotesize 
Fig. 2. - 
Left Panel: The predicted local color-magnitude relation: the color code represents the number density of galaxies, normalized to the maximum value.
Right Panels: The predicted color distribution  (differential number of objects per color bin, normalized to the total number) for galaxies of different magnitude (histograms) is compared with the data from the SDSS (Baldry et al. 2004, dots). 
\vspace{0.2cm}}

\begin{center}
\vspace{-0.4cm}
\scalebox{0.39}[0.39]{\rotatebox{-0}{\includegraphics{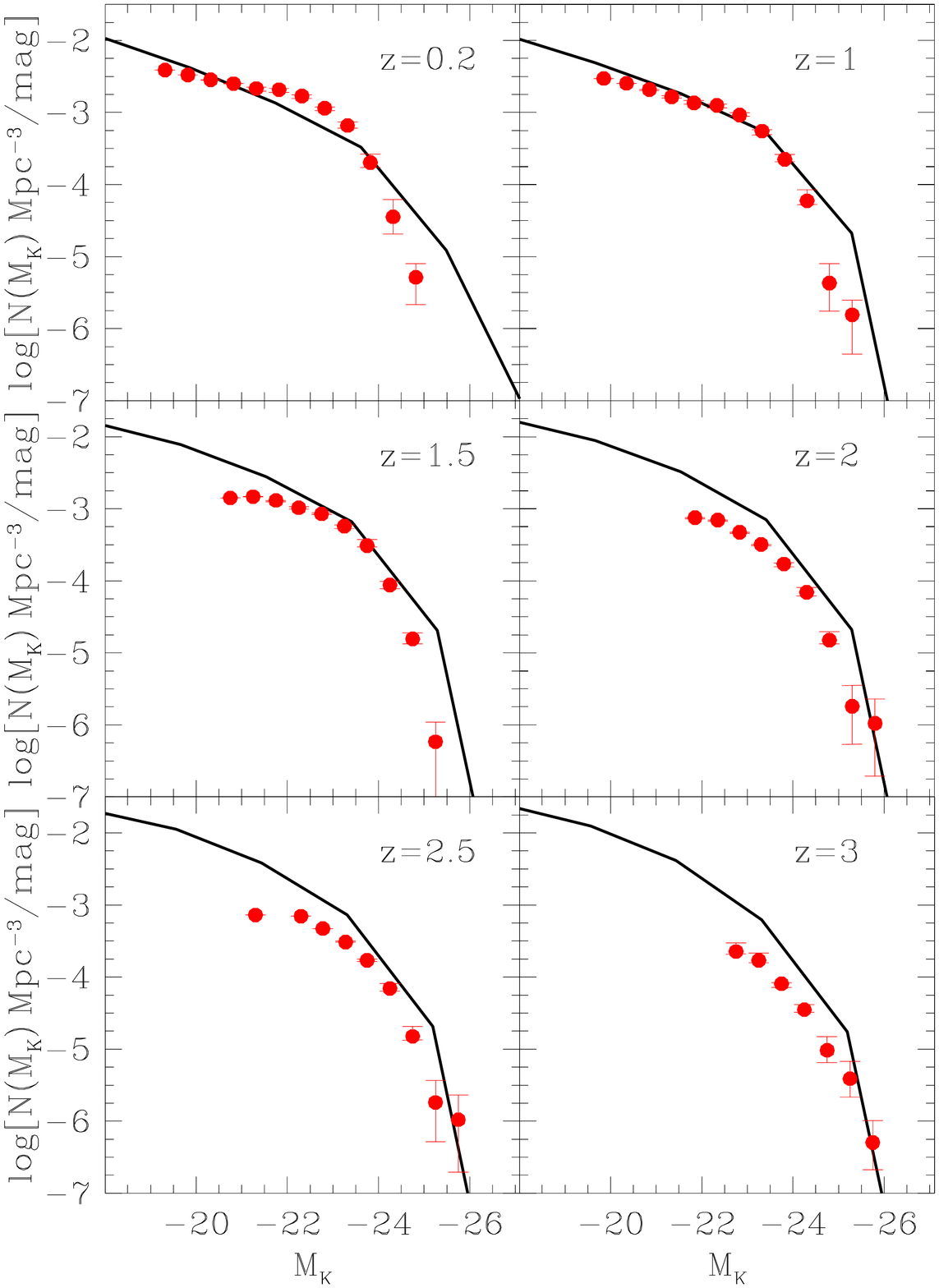}}}
\scalebox{0.397}[0.397]{\rotatebox{-0}{\includegraphics{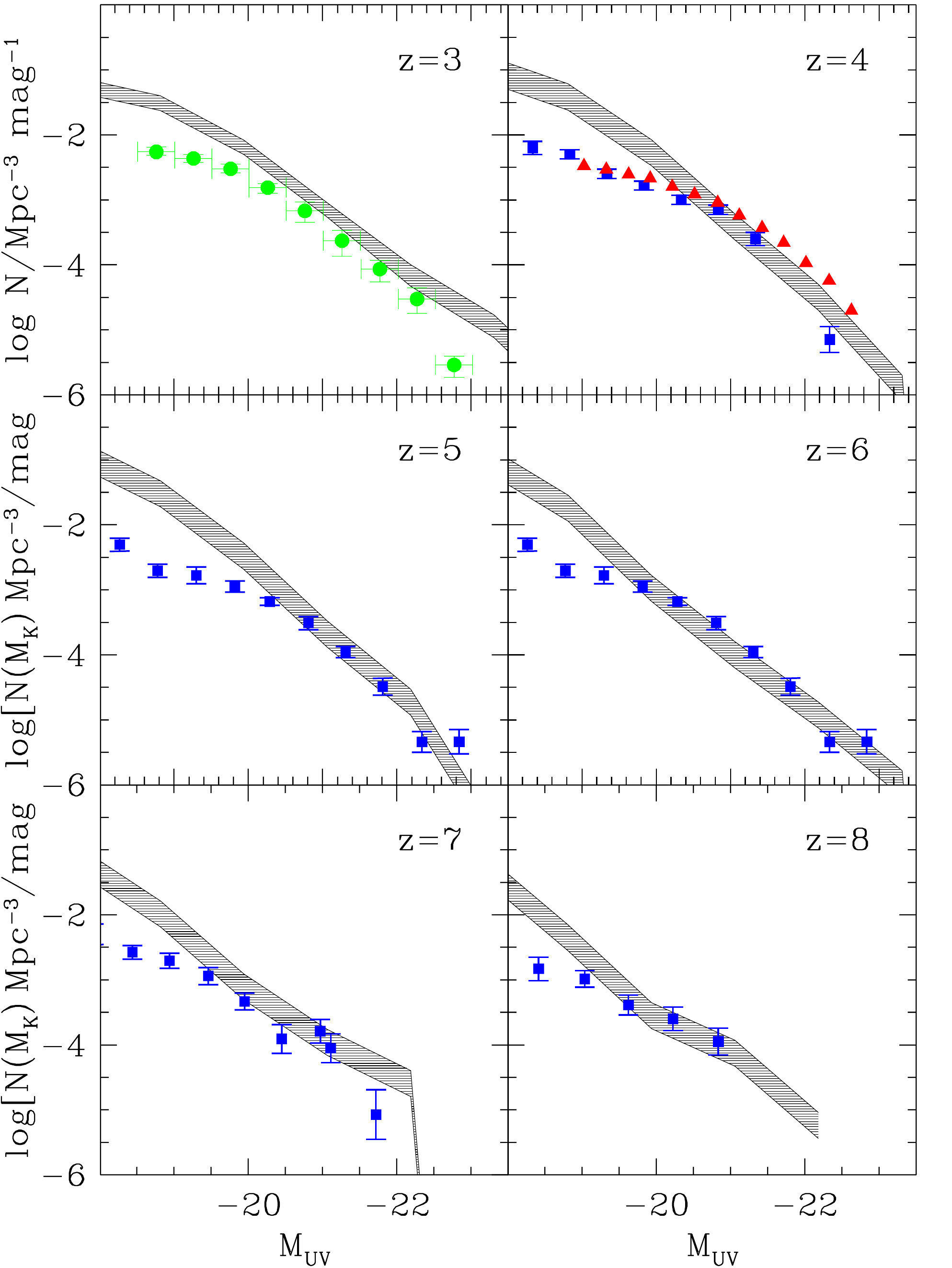}}}
\end{center}
\vspace{-0.3cm }
 {\footnotesize 
Fig. 3. - Left Panels: the evolution of the K-band luminosity function predicted by the model (solid line) is compared with the data from Cirasulo et al. (2010, squares). Right Panels: The predicted evolution of the UV luminosity function (shaded region) is compared with observed distribution of dropout galaxies. The shaded region encompasses the uncertainties due to the adopted extinction curve  (Small Magellanic Cloud, Milky Way or Calzetti) In each redshift bins, the model galaxies have been selected according to the same selection criterium adopted in the observations (see text). Data are taken from Reddy \& Steidel (2009, dots), Bowens et al. (2007, 2011, squares), and van der Burg, Hildebrandt, Erben (2010, triangles). 
\vspace{0.3cm}}

Finally we test that the evolution of the distribution of the $K$-band luminosity (a proxy for the stellar mass content) and of the UV luminosity (a proxy for the instantaneous star formation rate) 
are in reasonable agreement with the observations over the whole range of cosmic times where observations are available. This is done in detail in fig. 3, where (in the left panel) we compare our results with the observed evolution of the $K$-band luminosity function  from Cirasuolo et al. (2010) up to redshifts $\approx 3$ , and (in the right panel ) with the observed evolution of the UV luminosity functions of "dropout" galaxies from different authors (see caption) up to redshifts $z\approx 8$. Since the latter depends strongly on the color selection adopted to select the galaxy sample, we performed the same cuts in the color-color plane adopted in the observations to select $U$-dropout galaxies at $z\approx 3$ (Giavalisco et al. 2004) 
$B$-,$V-$, and $I$-dropouts (at $z\approx 4$, $z\approx 5$, $z\approx 6$, respectively, 
see Bouwens et al. 2007), and infra-red dropouts at $z\geq 7$ (Bouwens et al. 2011). To account 
for uncertainties related to adopted dust-extinction curve our predictions are shown as shaded regions (see caption). 

The comparisons above show that our semi-analytic model provides a reasonable statistical description of  basic physical quantities (such as star formation rates, assembled stellar mass) determining 
the cosmological evolution of the galaxy populations over a wide interval of cosmic times, so that it constitutes a solid "backbone" to connect the BH accretion (and hence the ensuing AGN evolution) to the physical properties of their host galaxies, as we describe next. 

\section{Implementing the AGN feeding processes}

We have updated our semi-analytic model as to include two different modes for the BH accretion: i) an interaction-triggered (IT) mode, where the triggering of the AGN activity is {\it external} and provided by galaxy interactions (including both mergers and fly-by), which 
can provide - for major interactions - a large relative variation $\Delta j/j$ of the angular momentum of the galactic gas; ii) a mode 
where accretion occurs due to disk instabilities, where the trigger is {\it internal} and provided by the break of the axial symmetry 
in the distribution of the galactic cold gas. In both cases, a nuclear star formation is associated to the accretion flow triggering the 
AGN. 

Our modelling for the IT mode has been described in previous papers 
(starting from Menci et al. 2004; 2008; see also Lamastra et al. 2013b), so we shall give here only a brief account of it, to recall its main properties. In this paper we expand on our implementation of the DI mode, which constitutes a novel component of the semi-analytic model;  it is based on the HQ11 analytical model; its detailed testing against high-resolution hydrodynamical N-body simulations (see HQ11; Angl\'es-Alc\'azar, \"Ozel, Dav\'e 2013) makes it a reliable baseline to describe the accretion flow onto the BH within the limits of the present modelling 
of instabilities. We shall recompute the HQ11 model, and extend it beyond the assumption originally adopted by HQ11 to asses its robustness, and to compute the {\it maximal} impact that the DI mode can have on the evolution of the AGN population under different hypothesis. 

\subsection{The Interaction-Driven Model}

As implemented in other semi-analytic models, galactic clumps within 
a host halo undergo different processes. First, they loose angular momentum due to dynamical friction and eventually merge with the central galaxy; such a process is implemented in our model
adopting the standard Chandrasekar expression for the dynamical friction timescale given in Lacey \& Cole (1993) adopted  in most semi-analytic models (see also Somerville \& Primack 1999; Cole et al. 2000; for variants of such a description see Somerville et al. 2012 and references therein). 

In addition, satellite galaxies with given circular velocity $v_c$ (and tidal radius $r_t$)  may interact inside a host halo with circular velocity $V$ at a rate
(Menci et al. 2002)
\begin{equation}\label{int}
\tau_r^{-1}=n_T(V)\,\Sigma (r_t,v_c,V)\,V_{rel} (V),
\end{equation}
where $n_T$ is the number density of galaxies in the host halo,
$V_{rel}$ the relative velocity between galaxies, and  the cross section for
such encounters. 
The above expression yields the probability for fly-by encounters (not leading to a merging) for  a geometrical cross section $\Sigma \sim \pi\langle r_t^2+r_t^{'2}\rangle$, while it determines the probability of a merging (binary aggregation) when the geometrical cross section is suppressed by a factor $\sim v_c^2/V_{rel}^2$ which accounts for the requirement that the orbital energy is transferred into internal degrees of freedom. The above appropriate cross sections are  given by Saslaw (1985) in terms of the tidal radius $r_t$ associated to a galaxy with given circular velocity $v_c$ (Menci et al. 2004). 
Note that the binary merging of satellite galaxies is also implemented in other semi-analytic models (see, e.g.,  Somerville \& Primack 1999), and that the tidal stripping of satellite galaxies is also included in our model as described in Menci et al. (2012).

The fraction $f$ of cold gas destabilized  by any kind of merging or  interaction has been worked out by  Cavaliere \& Vittorini (2000)  in terms of the  variation $\Delta j$ of the specific angular momentum $j\approx
GM/V_d$ of the gas to read (Menci et al. 2004):
\begin{equation}\label{fdest}
f\approx \frac{1}{2}\,
\Big|{\Delta j\over j}\Big|=
\frac{1}{2}\Big\langle {M'\over M}\,{R_d\over b}\,{V_d\over V_{rel}}\Big\rangle\, ,
\end{equation}
where $b$ is the impact parameter, evaluated as the greater of the radius $R_d$ and the average distance of the galaxies in the halo, $M'$ is the mass of the  partner galaxy in the
interaction, and the average runs over the probability of finding such a galaxy
in the same halo where the galaxy with mass $M$ is located.  The pre-factor accounts for the probability 1/2 of inflow rather than outflow related to the sign of $\Delta j$.
AGN and starburst events are triggered by all  galaxy interactions including  major mergers  ($ M \simeq M^{'} $),  minor
mergers ($ M \ll M^{'}$), and by  fly-by events.
We assume that in each interactions 1/4 of the destabilized  fraction $f$ feeds the central BH, while the remaining fraction feeds the circumnuclear starbursts  (Sanders \& Mirabel 1996). Thus, the BH accretion rate is given by 
\begin{equation}\label{macc_ID}
 {dM_{BH}\over dt}={1\over 4}{f\,M_c\over \tau_b} ~, 
\end{equation}
where the time scale $\tau_{b}=R_d/V_d$  is assumed to be the crossing time of the destabilized galactic disk.
Note that the fraction of destabilized gas in eq. \ref{fdest} is proportional to the mass ratio $M'/M$ of the merging partners; thus for grazing, low speed encounters  ($R_d\approx b$ and $V_{rel}\approx V_d$) major mergers (with $M'/M\approx 1$) may lead to 
a large relative variation of the gas angular momentum $\Delta  j\approx j$ and to large accreted fraction $f\sim 0.5$. 

\subsection{The Disk Instability Model}

In this Section, we a) recall the key points in the HQ11  computation of the inflow rate onto the BH, 	performed for a power-law gas density profile;  b) extend such a computation to treat different (exponential) gas density profiles; c) based on the HQ11 scenario, we work out the star formation rate associated to a given BH accretion rate; d) discuss how the computed accretion and star formation rates from disk instabilities actually represent upper limits to the real rates. In fact, in this work we are interested in investigating the {\it maximal} contribution that - according to present modelling - DI can give to the cosmological evolution of AGN, in order to constrain the evolution physical conditions where they may compete with, or even dominate, the contribution from 
galaxy interactions. 

\subsubsection{Computing the Black Hole accretion rate}
 
As recalled in the Introduction, gas rich mergers and tidal interactions are not the only way to produce a net mass inflow into the central region of the galaxies; in certain cases, secular growth processes such as disk instabilities (DI) might play a crucial role in feeding the central SMBH. 

From a modelling point of view, describing such an accretion mode requires i) modelling the DI trigger; ii) modelling the corresponding inflow rate onto the BH. As for the first, a widely used criterium for the onset of disk instabilities is 
that proposed by Efstathiou et al. (1982) based on N-body simulations, which states that a disk becomes unstable if its mass exceeds a critical value:
\begin{equation}
M_{crit} =  {v_{max}^2 R_{d}\over G \epsilon}
\end{equation} 
where $R_d$ the scale length of the disk, and $v_{max}$ is the maximum circular velocity considering all the galaxy components; the latter is derived from velocity profiles computed assuming a NFW dark matter density profile (Navarro et al. 1997), and including the effect of baryons after Mo, Mao, White (1998), as described in Menci et al. (2005).  Here $\epsilon\sim 0.5 - 0.75$ is a parameter calibrated on simulations. The above criterium is that usually adopted in semi-analytic models: Hirschmann et al. (2012) adopt a value $\epsilon = 0.75$, and a similar value is adopted by Fanidakis et al. (2011). In order to investigate the maximal affect of DI 
on the statistical evolution of AGN we also adopt the value $\epsilon = 0.75$.

Computing the mass inflow rate onto the BH, and the associated nuclear star bursts, is a more complex issue. At present, semi-analytical model only resort to {\it ansatz} tuned to maximize the model agreement with observations; in Hirschman et al. (2012) 
the mass accretion rate is then calculated considering all the mass exceeding the threshold to be added in the central bulge and a tiny fraction of it ($\sim 10^{-3}$, motivated by the local BH-bulge mass relation) to be accreted in the central BH according to an exponential growth characterized by the Salpeter timescale. In the GALFORM model Bower et al. (2006) and Fanidakis et al. (2012) assume that all the mass of the disk (not only the mass in excess) is added to the central bulge, leading to an enhanced effectiveness of the whole process.
Due to the lack of an analytical description of the complex physics processes acting in the transport of gas in the nuclear region, the results strongly depend on the assumed prescriptions and may actually overestimate the real inflows attainable in the central regions under realistic conditions.

Here we attempt to adopt a more physical treatment of the mass inflow from disk instabilities. In particular,  we take on the analytical model proposed by HQ11, which aims at accounting for the physics of angular momentum transport and gas inflow from galactic scales ($\approx 1$ kpc) inward where the BH potential is dominant. It considers the multi-component nature of the disk, and the inflow toward the BH is related to the nuclear star formation. In addition, such a model has been tested by the authors against aimed hydrodynamical simulations, so that at present it constitutes a solid baseline to describe the BH accretion due to disk instabilities. 
 
The model investigates the response of the disk to an external non-axisymmetric perturbation of the potential $\Phi_a = |a|\Phi_0$, where the perturbation amplitude is $|a| < 1 $; the unperturbed system corresponds to an axisymmetric thin disk with a spherical component (BH, bulge and dark matter halo). While for periodic particle orbits the net angular momentum loss is null at the first order in $a$, HQ11 showed that the shocks generated by gas particle in the condition of orbit crossing lead to a first-order effect; as a result, in the orbit crossing (non-linear) regime they obtained the orbit-averaged gas inflow rate 
\begin{equation}
\label{massinflow}
\frac{dM_{infl}}{dt} = \Sigma_{g}(R)\,R^2 {\Omega\over 2} \frac{\Phi_a}{V_c^2} \frac{m}{1+ \partial ln V_c / \partial ln R} 
\end{equation}
where $\Sigma_g(R)$ is the gas surface density profile, $\Omega$ is the angular velocity of the unperturbed disk, induced by both the bulge and the disk mass enclosed within the radius $R$, and $V_c$ is the corresponding circular velocity,  $m$ is the mode of the perturbations (the perturbed potential is $m$-fold symmetric, so that $m=0$ corresponds to axial symmetry), and marginal orbit crossing is assumed following HQ11 . Here, 
our strategy is to {\it maximize} the effects of DI to single out the wider range of physical conditions where such a process may be effective in fueling AGNs. Hence,  we will assume that when disk instabilities are triggered the above eq. \ref{massinflow} always holds. 

The mass inflow rate in eq. \ref{massinflow} is of first order in perturbation amplitude, since it depends linearly on $\Phi_a$.
In order to calculate the exact value of the mass inflow rate, we need the analytic expression for  $\Phi_a$ and $\Sigma_{g}$. The former is obtained in the WKB approximation where the disk dominates the potential, while it is related to the BH mass in the inner region 
where the BH gravity is dominant (see Appendix A);  the latter can be estimated assuming an equilibrium between mass inflow rate due to the disk instability and star formation rate (described in terms of its surface density $\Sigma_*$): 		

\begin{equation}
\label{masstot}
{\partial\over \partial R} {dM_{infl}\over dt} \simeq 2\,\pi R\,{d\Sigma_*\over dt}=2\,\pi\,R\,{\Sigma_k\over t_k}\,\Bigg({\Sigma_g\over \Sigma_k}\Bigg)^{\eta_k}
\end{equation}

The last equality follows after relating the star formation surface density $\Sigma_*$ to the gas surface density $\Sigma_g$ 
through the Schmidt-Kennicut law $d\Sigma_*/dt=(\Sigma_k/t_k)*(\Sigma_g/\Sigma_k)^{\eta_k}$, where the timescale $t_k$, the normalization $\Sigma_k$ and the exponent $\eta_k$ are derived from fits to the observations: typical values (Bouche' et al. 2007) 
are $t_k\approx 0.4\,10^9$ yr, $\Sigma_k\approx 10^8$ M$_{\odot}$ kpc$^{-2}$ and  $\eta_k=1.7$, but the real slope $\eta_k$ is still subject to different observational estimates in the range  $1.2\lesssim \eta_k\lesssim 1.8$ 
(see, e.g. Davies et al. 2007; Hicks et al. 2009; Daddi et al. 2010; Genzel et al. 2010;  Santini et al. 2014), also depending on the redshift and on the properties of the adopted galaxy sample. 

There could be situations when the equilibrium described by eq. \ref{masstot} does not hold; when star formation inhibits completely the mass inflow, or when the gas inflow rate is larger than star formation  (Barnes $\&$ Hernquist, 1991). However, we later show that these two different regimes yield a mass inflow onto the central BH weaker than inflow obtained when the equilibrium is reached.
Since our aim is to study the {\it maximal} effects of DI on the cosmological evolution of AGN we shall retain the above approach. 

Using eqs. \ref{massinflow} and \ref{masstot} it is possible to solve for both the gas density (and hence for the star formation) and 
the mass inflow. To this aim (see Appendix A for a step-by-step computation), the  potential (and hence $\Phi_a$ in eq. \ref{massinflow}) has to be evaluated in two regions: an outer region where the disk gravity is dominant, and an inner region where gravitational potential is dominated by the BH.  The radius $R_{BH}\sim $ 10-100 pc (depending on the BH mass) marks the separations between the two regions, while the scale $R_{acc}\sim 10^{-2}-10^{-1}$ pc corresponds to the inner bound of the innermost regions; inside this radius star formation ceases and the mass inflow can be assumed to be almost constant. 
The mass inflow at this radius $\dot M_{infl}(R_{acc})$ corresponds to the final BH accretion rate that we shall include in our semi-analytic model;  the computation of such a quantity is given in detail in Appendix A, where it is expressed in terms of quantities 
evaluated far from the very central region, at a "galactic" scale $R_0\sim 100-1000$ pc, so that the central BH accretion rate can be easily connected to the galactic properties of the host galaxy computed in our semi-analytic model. This is done in Appendix A for a power-law scaling of the disk surface density $\Sigma_d\sim R^{-\eta_d}$ and in Appendix B for an exponential  disk profile. In the first case, $\eta_d$ is calibrated on N-body simulations:  our fiducial choice is to take the best fitting value $\eta_d=1/2$ given by HQ11. Then, for a Kennicut-Schmidt 
index $\eta_k=7/4$  (as adopted by HQ11) the final result is 

\begin{equation}
\label{hopkins}
\frac{dM_{BH}}{dt}  \equiv \frac{dM_{infl}(R_{acc})}{dt}    \approx 
{\alpha(\eta_k,\eta_d) \,
f_d^{4/3}\over 1+2.5\,f_d^{-4/3}(1 + f_0/f_{gas}) }
\left( \frac{M_{BH}}{10^8 M_{\bigodot}}\right)^{1/6}
\left( \frac{M_d(R_0)}{10^9 M_{\bigodot}}\right)
\left( \frac{R_0}{100 pc }\right) ^{-3/2}
\left( \frac{R_{acc}}{10^{-2}R_{BH}}\right) ^{5/6}
M_{\bigodot} yr^{-1}
\end{equation} 
\begin{equation}
f_0 \approx 0.2 f_d^2 \left[ \frac{M_d(R_0)}{10^9 M_{\bigodot}}\right]^{-1/3}
{\hspace{2cm}}  f_{gas} \equiv {M_{gas} (R_0)\over M_d(R_0)}
\end{equation} 
 
Here $M_{BH}$ is the central black hole mass,  $f_d$ is total disk mass fraction, and  $M_d$ and $M_{gas}$ are the the disk and the gas mass calculated in $R_0$, respectively. For physical values of $0\leq f_d\lesssim 0.5$ the above equation yields accretion rates indistinguishable from the original formulation given in HQ11. The above equation relates the BH accretion at the sub-resolution scale $R_{acc}$ to quantities evaluated at the galactic scale $R_0$ which are computed for each galaxy by the semi-analytic model described in sect. 2. Note that - under the assumptions adopted by HQ11  - the product $M_d(R_0)\,R_0^{-3/2}$ is constant (as shown in our Appendix A), so that eq. \ref{hopkins} can be evaluated at any radius $R_0$, provided it remains within the range where the adopted assumptions for the scaling of the stellar disk surface density remain valid ($R_0\lesssim 1$ kpc). 
 The constant $\alpha(\eta_k,\eta_d) \sim 0.1-5 $ is given in eq. \ref{alpha} and parameterizes the strong dependence of the normalization on the exact value of $\eta_k$ of the Schmidt-Kennicutt law (see Appendix A). In order to {\it maximize} the contribution of the DI mode to the evolution of AGN, in the  following we shall keep such a parameter to a maximum value $\alpha(\eta_k,\eta_d)=10$. Finally, following HQ11, we assume $R_{acc}$ to be equal to $ \sim 10^{-2} R_{BH}$. 

Since our semi-analytical model uses exponential-law density profiles to describe the disk surface density profile (at galactic scales), we have developed all the calculations also for the case of $"$exponential disks$"$. The resulting mass inflow rate is the following

\begin{equation}
\label{exponential}
\frac{dM_{BH}}{dt} \approx \alpha(\eta_k) \,f_d^{19/12} 
\left( \frac{M_{BH}}{10^8 M_{\bigodot}}\right)^{5/12}
\left( \frac{M_d(R_0)}{10^9 M_{\bigodot}}\right)^{3/4}
\left( \frac{R_0}{100 pc }\right) ^{-3/2}
\left( \frac{R_{acc}}{10^{-2}R_{BH}}\right) ^{5/6} M_{\bigodot} yr^{-1}
\end{equation}

with $R_d$ the exponential length of the disk; all the calculations for exponential disks are shown in Appendix B. All the factors in eq. \ref{exponential} are the same of eq. \ref{hopkins}; $\alpha(\eta_k,\eta_d)=0.1-1.2$ 
(depending on $\eta_k$) is now given by eq. \ref{alpha} with $\eta_d=0$. 
With the choice $R_{acc} \sim 10^{-2}R_{BH}$, the inflow is quite insensitive to the exact values of $R_{acc}$ and $R_{BH}$; however, when it is necessary (see next section), we will eventually take $R_{acc} \sim 0.1$ pc and $R_{BH} \sim 10$ pc. 

\subsubsection{Computing the nuclear star formation rate}

Since the above models assume equilibrium between mass inflow and star formation, a star formation activity is always connected to the 
BH acccretion rate given in eq. \ref{hopkins} or \ref{exponential}, and can be computed in detail after eq. \ref{masstot}, as shown in Appendix A. 
As a result, a star formation rate $\dot M_{*,DI}= A_{*,DI}\dot M_{BH}$ is obtained, where the exact value of the proportionality constant 
$A_{*,DI}\gtrsim 10^2$ depends on the detailed radial profile of the disk potential and of the gas surface density. The effects of changing $A_{*,DI}$ 
within the limits derived in Appendix A  and B will be quantitatively shown in the next section; here we note that, given a total amount of baryons available for accretion, increasing $A_{*,DI}$ leads to a faster gas consumption, and hence to a lower accretion rates at subsequent times, with the effect of decreasing the number of active AGNs. Thus, to investigate the {\it maximal} impact of the DI mode on the growth of BHs and on the evolution of AGNs
it is sufficient to adopt  proper lower limits for the DI star formation rate. This can be done within the framework of the DI model introduced above, and the derivation is shown (Appendix A, B),  for both a power-law  or and an exponential profile of the disk surface density. We obtain 
a lower limit $A_{*,DI}=100$ (eq. A18) for a power-law profile (see Section A.5), corresponding to the contribution to the star formation 
from the inner region $R\lesssim R_{BH}$; this also constitutes a lower limit  for the case of an exponential disk, since in the inner region the potential 
determining the mass inflow is dominated by the BH and does not depend on the assumed disk profile (see also Appendix B, eq. B11-B13). The contribution to the star formation rate from outer regions depends strongly on the assumed disk surface density; in the case of a power-law profile, 
larger and larger contributions are obtained when the power-law behaviour is extrapolated  to regions far from the nucleus (see eq. 
A19-A22), while in the case of an exponential profile (more realistic in regions far from the BH sphere of influence $R\gg R_{BH}$) we obtain an additional contribution saturating to $A_{*,DI}=60$. 

\section{Results}

We compute separately the effects of the two accretion modes on the AGN population. To compare with observations, the AGN bolometric luminosity is computed as 
\begin{equation}\label{Lagn}
L_{AGN}=\eta\,c^2\,{d M_{BH}\over dt}
\end{equation}
where $dM_{BH}/dt$ is taken from eq. \ref{macc_ID} for the IT mode, and from eq. \ref{hopkins} (or eq. \ref{exponential}) for the DI models. We
we adopt an energy-conversion efficiency $\eta= 0.1$ (see Yu \&
Tremaine 2002). The luminosities in the UV and in the X-ray bands are computed from the above expression 
using the bolometric correction given in Marconi et al. (2004). 

We stress again that in the present work we do not attempt to find a best-fitting combination of the two accretion modes. Rather, we are interested in isolating the effects of the two accretion modes on the evolution of the AGN population, to single out the physical regimes where they may be effective in driving the AGN evolution. Given that in our previous paper (and in the following results) 
the IT mode has been shown to be able to provide acceptable fits to the observed evolution of the AGN population, 
our strategy will be to {\it maximize} the possible DI accretion rates in order to assess if and for what kind of objects we expect it to be competitive with the IT mode. To this aim, we shall explore the possible variants of the DI model provided by different choices for 
i) the normalization $\alpha(\eta_K,\eta_d)$ in equation \ref{hopkins}, which in turn is related to the present uncertainties concerning the slope of the Kennicut-Schimdt law $\eta_k$; ii) the different disk surface density profiles, determining the slope $\eta_d$ (see Sect. 3.2.1) and the dependence of $\dot M_{BH}$ on the disk mass fraction and on the gas fraction (through the boundary condition $A(R_{BH})$  in Appendix A; iii) the star formation associated to the DI mode and discussed in Sect. 3.2.2.  

Such quantities are not freely tunable, since they can only vary within limits set by observations and simulations. The former set a limit for the Kennicut-Schmidt index $3/2\leq \eta_k\leq 7/4$ (see sect. 3.2.1). The latter provide a range of possible  gas surface density profiles ranging from $\Sigma_g\sim R^{-3}$ to $\Sigma_g\sim R^{-1}$,  bracketing the effects of the present uncertainties in the physics of the inter-stellar medium (see Hopkins \& Quataert 2010); these concern mainly the effective equation of state of the gas, which widely varies from cases where it is quite stiff on small scales (near-adiabatic, similar to what is adopted in the studies of Mayer et al. 2007; Dotti et al. 2009), through to cases where the gas is allowed to cool to a cold isothermal floor and forms a clumpy, inhomogeneous medium (on galactic scales, similar to what is assumed in Bournaud et al. 2007; Teyssier et al. 2010), and includes cases where the gas motions gain a significant contribution from resolved turbulent motions and the gas mass distribution can be highly inhomogeneous/clumpy. The above range of uncertainty for the gas surface density profiles translates (eqs. A1, A2, B8 ) into values ranging from $\eta_d=1/2$ (the HQ11 case) to $\eta_d=0$ (our exponential case).  

For the above ranges of uncertainty $3/2\leq \eta_k\leq 7/4$ and $0\leq \eta_d\leq 1/2$, the normalization of 
the BH accretion rates in eq. 8 and 9 remains in the range $0.1\lesssim \alpha(\eta_k,\eta_d)\leq 5$ (see sect. 3.2.1). To investigate the {\it maximal} impact of DI we adopt a safe  normalization $\alpha(\eta_k,\eta_d)=10$, the same considered by HQ11 to constitute a definite upper limit. A final source of uncertainty comes from the the contribution to the nuclear star formation rate from regions outside the BH sphere of influence $R\geq R_{BH}$; assuming it to be null sets a lower limit $A_{*,DI}=100$ for the normalization of the nuclear star formation rate, which maximizes the amount of inflowing gas feeding the  BH accretion, as discussed in sect. 3.2.2. We shall refer to the combination  $\alpha=10$ and $A_{*,DI}=100$ as "maximal" DI model. In addition,  we shall also explore the effect of the additional contribution to the nuclear starbursts from regions far from the BH sphere of influence. 

In sum, following the lines above, we  explore three variants of the DI mode: \\
$\bullet$  a {\it fiducial case} corresponding to the model adopted by HQ11 (eq. \ref{hopkins}), computed  assuming a power-law disk surface density profile $\eta_d=1/2$ with a {\it maximized} value of the parameter $\alpha(\eta_K,\eta_d)=10$ (see Sect. 3.2.1), and assuming the lower limit for the associated star formation rate $A_{*,DI}=100$ (see Sect. 3.2.2  and Appendix A). \\
$\bullet$ the case of an {\it exponential profile} for the disk surface density (eq. \ref{exponential}), keeping the same values for $\alpha(\eta_K,\eta_d)=10$ and $A_{*,DI}=100$ (see Sect. 3.2.2  and Appendix B). \\
$\bullet$ a model with {\it enhanced star formation} rate associated to DI (for a power-law profile); it assumes a normalization of the star formation associated to DI above the lower limits derived in Sect. 3.2.2, with an additional contribution $A_{*,DI}=100$ from regions far from the BH sphere of influence ($R\gg R_{BH}$). 
 

Having defined our reference set of DI cases we proceed to study the evolution of the AGN population under
different triggering models.s

\subsection{The Evolution of the AGN luminosity function and BH mass function}

The evolution of the AGN UV luminosity funtions predicted in the IT scenario is compared with that corresponding to the DI mode  in fig. 4. The luminosity functions are computed in the UV band at $\lambda = 1450 {\AA}$, where most of the UV rest frame data on AGN are collected and where the average AGN emission shows the peak in the spectral energy distribution, representative of the overall bolometric emission. The predictions are compared with data taken from different bands, including the X-ray data (corrected for obscuration) by La Franca et al. 2005; Ebrero et al. 2009 at low redshifts $z\lesssim 1.5$, the X-ray data for objects in the COSMOS field by Brusa et al. 2010;  Civano et al. 2011; and the {\it Chandra} X-ray data for objects detected at $z\geq 4$ in the GOODS south field by Fiore et al. 2012); these are converted to the UV with the bolometric corrections in Marconi et al. (2004). Details on the conversion procedure are given in Fiore et al. (2012) and Giallongo et al. (2012). 

\begin{center}
\vspace{-0.cm}
\scalebox{0.57}[0.57]{\rotatebox{0}{\includegraphics{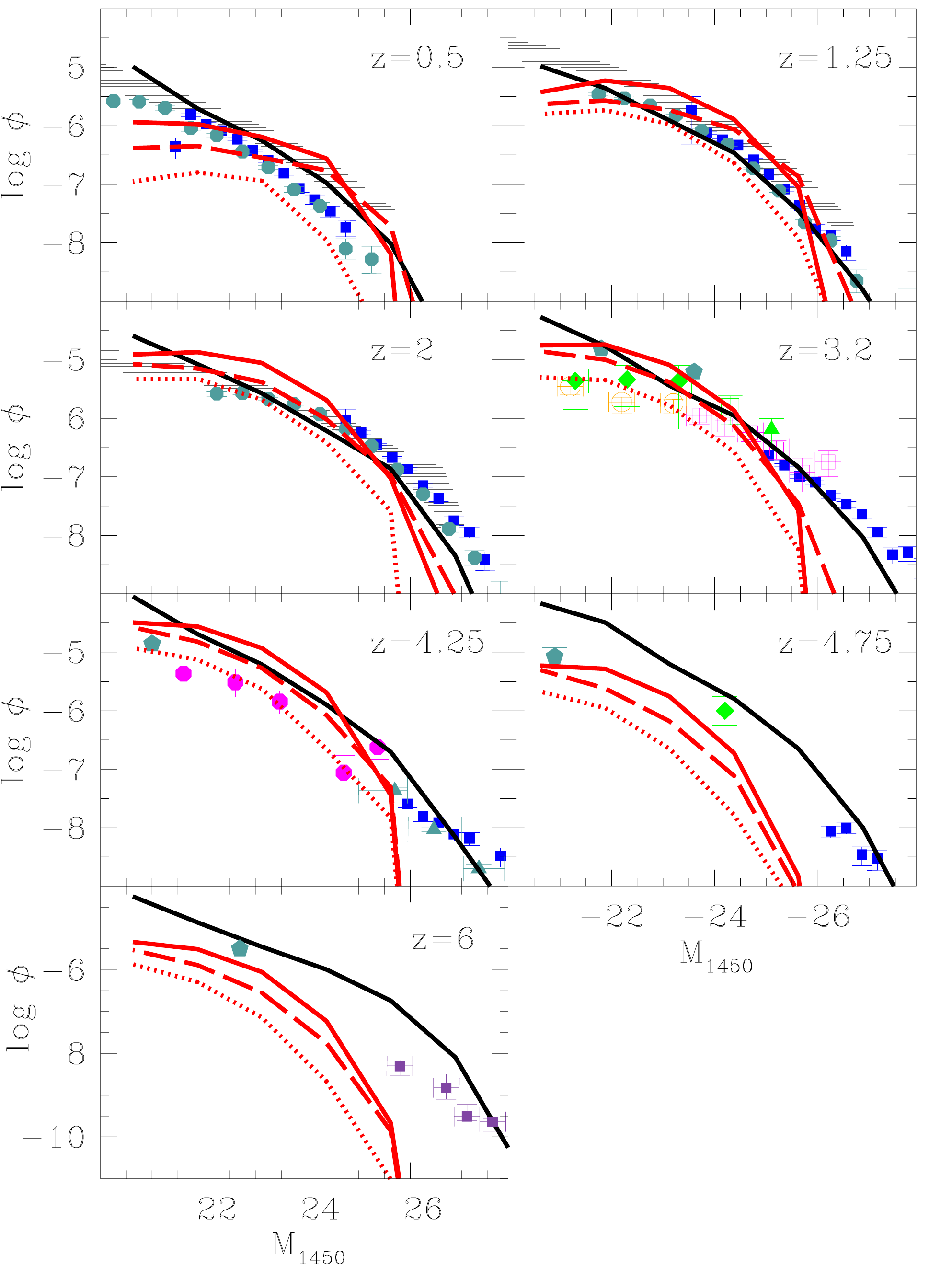}}}
\end{center}
\vspace{-0.5cm }
 {\footnotesize 
Fig. 4. - The predicted evolution of the AGN luminosity function in the UV band (at 1450  ${\AA}$). Solid black lines refer to the  IT scenario, while  red solid lines to the DI model. We also plot variants of the DI model: dashed lines refer to the DI model computed with an exponential profile, while dotted lines to the DI model with increased star formation (see text). 
We compare with observed UV luminosity functions by  Richards et al. (2006, solid squares), Croom et al. (2009, solid circles at $z<3$ ), Siana et al. (2008, open squares), Bongiorno, Zamorani,  Gavignaud (2007, open circles), Fontantot et al. (2007, solid diamonds), Glikman et al. (2011, big filled circles at $z=4.25$), Jiang et al. (2009; solid squares at $z\geq 6$). We also include luminosity functions from X-ray observations, converted to UV through the bolometric corrections in Marconi et al. (2004):  the shaded region brackets the determinations derived in the X-ray band at $z\leq 2$ 
by La Franca et al. (2005), Ebrero et al. (2009), Aird et al. (2010), while at higher redshifts we plot data from  Fiore et al. (2012, penthagons), Brusa , Civano, Comastri (2010, triangle), Civano, Brusa, Comastri (2011, green empty squares), . 
\vspace{0.cm}}

The results shown in fig. 4 indicate that none of the variants of the DI mode can  provide the accretion needed to obtain high-luminosity QSO with $M_{1450}\leq -26$, especially at redshifts $z\gtrsim 3.5$; this is because the accretion rates in eqs. \ref{hopkins} and \ref{exponential} have a mild dependence on the BH mass, but depend very strongly on the disk mass fraction and on the gas mass fractions ($f_d$ and $f_{gas}$). Thus, in galaxies with relatively low gas fraction, or in galaxies with a large B/T ratio, the DI mode is more suppressed than the IT mode. As expected DI can be instead very effective in the range of galaxy masses (and AGN luminosities) dominated by disk galaxies $-25\lesssim  M_{1450}\lesssim -22$; while we must keep in mind that the predicted DI luminosity functions in fig. 4 constitute {\it upper limits}, the results show that in such a magnitude range the DI accretion rates could compete with - or even overtake - those attained in the IT scenario, at least for redshifts $ z\lesssim 3.5$. At higher redshifts, although galaxies are surely gas rich with large $f_{gas}\sim 1$, the typical disk mass fraction $f_d$ decreases since - in the standard scenario assumed in semi-analytic models - the frequent major interactions continuously disrupt the forming disks. 

Note that - within the assumed modelling for the inflow driven by DI - our fiducial DI case constitutes an effective upper limit for the bright-end of the luminosity function; increasing the associated star formation $A_{*,DI}$ above the lower limit $A_{*,DI}=100$ would yield lower luminosity functions, as shown by the dot-dashed line in fig. 4, for the reasons explained in Sect. 3.2.2. Changing the disk power-law density profile assumed in the original version of HQ11 to an exponential profile also 
results in a lower abundance of bright AGN. Indeed, the same fiducial model must be regarded as an effective upper limit, since the assumed value for the normalization $\alpha(\eta_K,\eta_d)=10$ is actually appreciably larger than any value attained with realistic values of $\eta_k$ and $\eta_d$, as shown in Appendix A (see eq. A14 and the text below). 

For the IT model, the results presented here are very similar to those obtained in the earlier version of our model (see, e.g., Giallongo et al. 2012), 
and show an acceptable matching with the data. For $z\geq 4.5$ the real size of the model over-prediction with respect to the faintest data points 
is still to be assessed. In fact, the observations by Fiore et al. (2012)  could be affected by several uncertainties and significant incompleteness; the objects have been selected from very deep Hubble Space Telescope NIR images at $H \approx 27$ and measured at the faintest X-ray fluxes of $F_X \approx 2 × 10^{−17}$  erg s$^{-1}$ cm$^{-2}$. At the lowest X-ray luminosities, objects with relatively high X/optical ratio, and consequently with $H > 27$, are missed in their survey.  As a consequence, uncertainties in the Fiore et al. (2012) luminosity functions  data due to systematic errors could be larger than shown in the figure based on number statistics. 

While the above result strongly disfavor DIs as the dominant accretion mode in high-luminosity AGNs, they also indicate that it could be as effective as the IT mode (or even more) in intermediate-luminosity AGNs especially at intermediate redshifts $2\lesssim z\lesssim 3.5$; this is the epoch where massive star formation - possibly triggered by DI in gas-rich objects - is observed in disk galaxies (Bournaud et al. 2007; Genzel et al. 2008). Thus, we can seek for proper observables able to assess whether one of the two modes clearly dominates in this luminosity and redshift range. 

A possible approach could be to seek for signatures in the BH mass function and/or in the $M_{BH}-M_*$ relation. 
The former describes statistically the growth of BH masses corresponding to the evolution of the BH accretion rates shown in fig. 4. Our predictions for such a quantity are presented for both the IT and the fiducial DI models in fig. 5. At low redshifts the predictions are compared with a compilation of data (Shankar, Weinberg, Miralda-Escud\'e 2009). While present data do not allow to discriminate between the DI and the IT scenario at low redshift $z\lesssim 0.3$, the predicted evolution of the BH mass funtions differs appreciably at higher redshifts; in particular, as expected  from the evolution of the AGN luminosity functions, DI provide a much lower abundance of massive BH at $z\gtrsim 1$. 

\begin{center}
\vspace{-0.2cm}
\scalebox{0.45}[0.45]{\rotatebox{-90}{\includegraphics{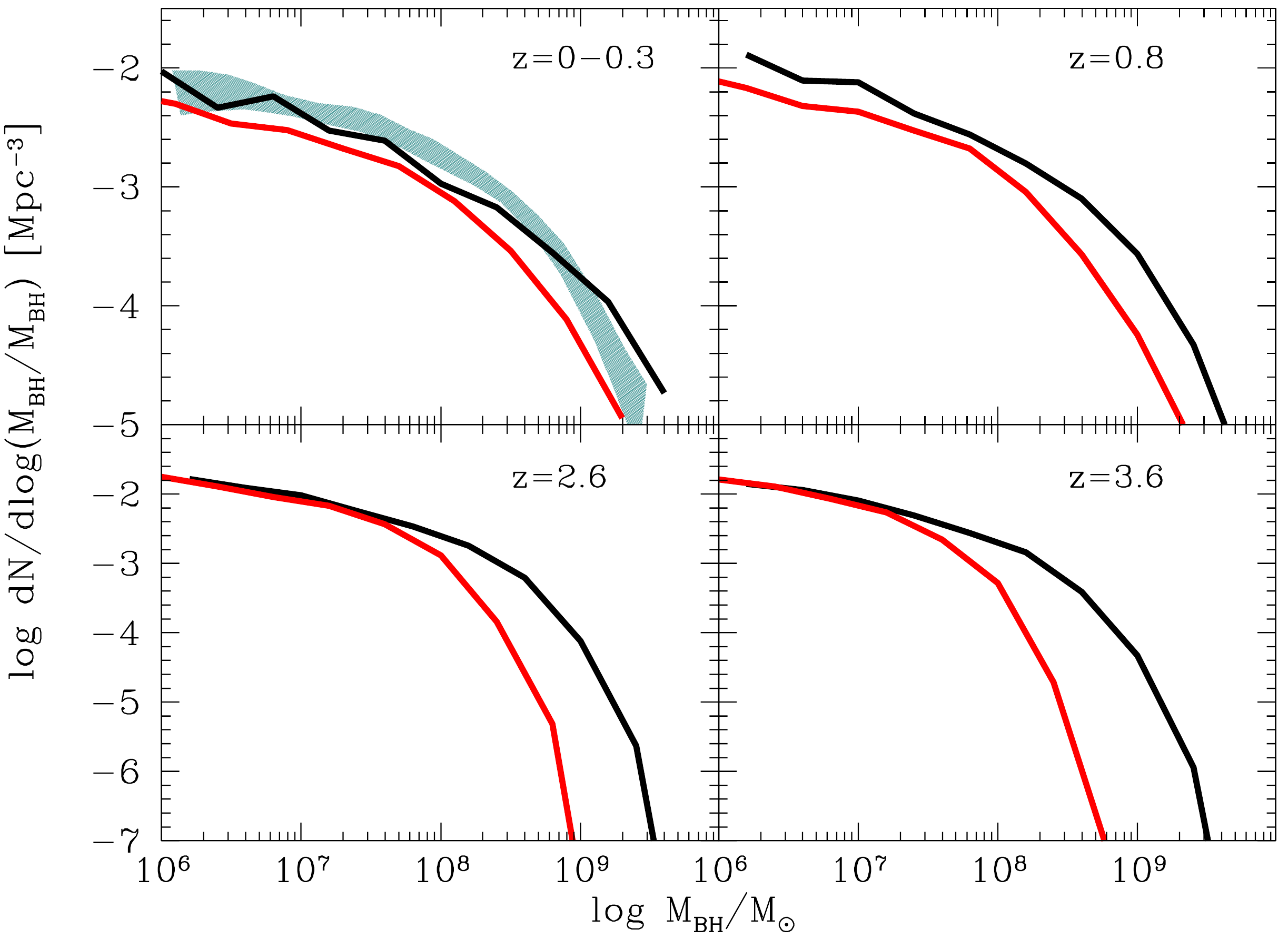}}}
\end{center}
\vspace{-0.cm }
 {\footnotesize 
Fig. 5. - The predicted evolution of the Black Hole mass function in the Interaction-Driven scenario (solid black lines) and in the Disk Instability model (solid red line).
The shaded region defines the spread in observational estimates obtained using different methods, as compiled by Shankar, Weinberg, Miralda-Escud ́e (2009).
\vspace{0.1cm}}

\subsection{The $M_{BH}-M_*$ relation}

A larger difference between the IT and the DI prediction is shown by the  local $M_{BH}-M_*$ relation, shown in fig. 6. While both the IT and the DI accretion modes provide a local relation consistent with observations, the slope and - most of all - the scatter differ appreciably.
While present data do not allow to clearly favour one of the two modes, it is interesting to note that the large scatter characteristic of the IT mode originates from the variance in the merging histories, which has a large impact in a model where interactions provide the trigger and the power of the AGN emission; the origin of such a variance can be visualized through the paths followed by AGN host to reach their final position in the $M_{BH}-M_*$ plane. 

Such paths include situations in which the progenitor of present-day massive BH are characterized by an early accelerated phase of BH growth which brings them {\it above} the local relation at early times $z\gtrsim 4$, followed by a later phase where the BH growth appreciably decelerates while the "quiescent" star formation 
gradually increases the stellar content of their host galaxies; these paths are those leading to the early building up of large-mass BHs at high redshifts $z\gtrsim 4$. The paths passing below the local relation, instead, correspond to galaxies retaining a large fraction of their gas down to late cosmic epoch, so that they form stars at very high rates down to $z\approx 2$ (the most star forming objects are those corresponding to SMG, see Lamastra et al. 2010). Such a large variance in the paths leading to the local relation is missing in the DI scenario, where the trigger and the accretion rate of AGN are provided by internal properties, less prone to the merging history. Indeed, in the DI the paths are confined in a small region of the plane; the growth of BH follows a   corresponding growth of the stellar mass content, and phases where either $\dot M_{BH}$ or $\dot M_*$ largely dominate are absent. This means that we expect a tight correlation between the AGN luminosity and the star formation rate in the host galaxy. This constitutes an interesting point that we shall 
investigate in detail in a following paper. 

\begin{center}
\vspace{0.cm}
\scalebox{0.4}[0.4]{\rotatebox{0}{\includegraphics{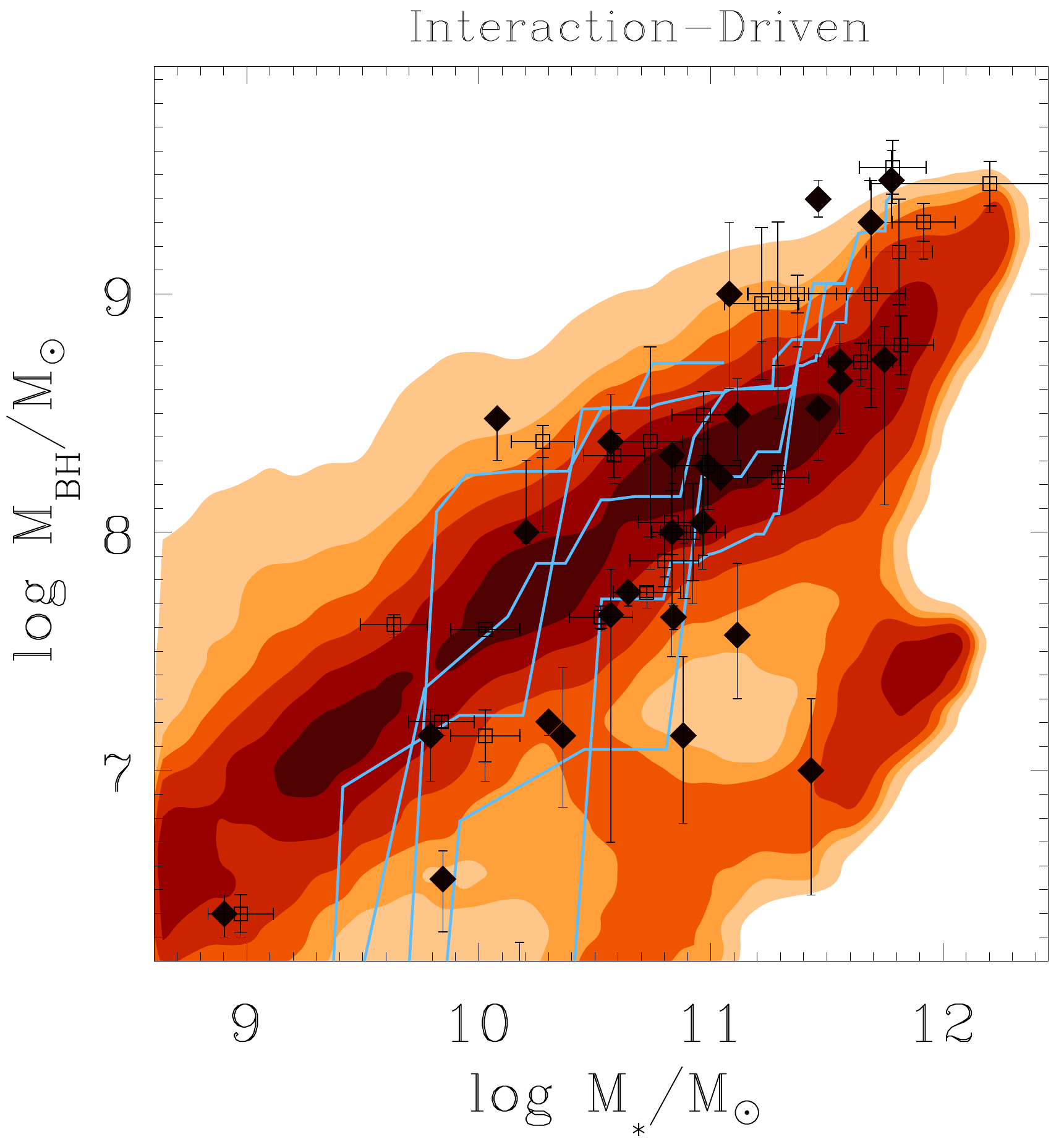}}}
\hspace{1cm}
\scalebox{0.4}[0.4]{\rotatebox{0}{\includegraphics{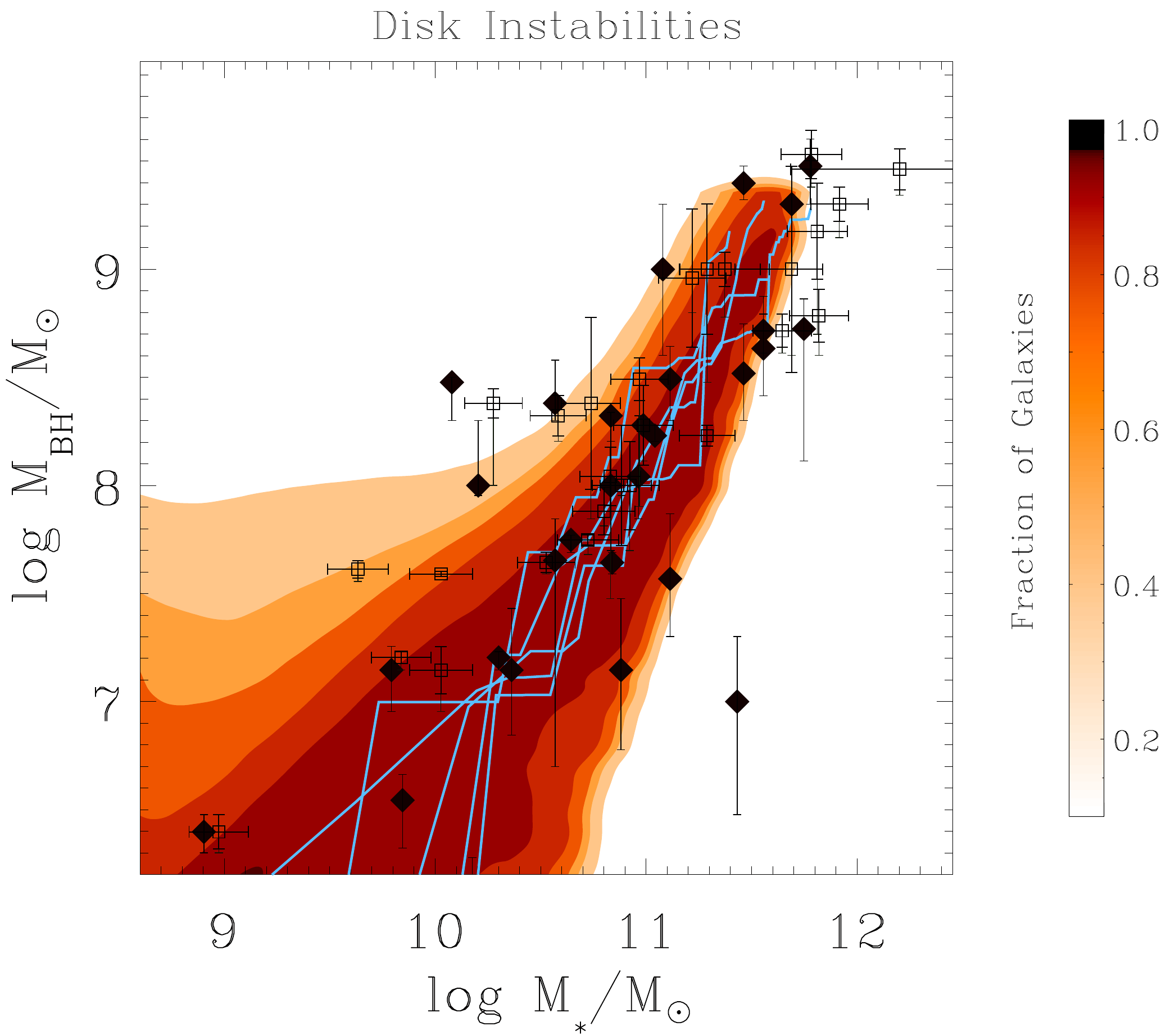}}}
\end{center}
\vspace{-0.2cm }
 {\footnotesize 
Fig. 6. -  The local $M_{BH}-M_*$ relation  for the Interaction-Driven model (left panels) 
 and the Disk-Instability model (right panels). 
Data points represent the observed local relation from  H\"aring and Rix (2004, diamonds), and 
Marconi \& Hunt (2003, squares, here $M_*$ is derived using the best-fitting virial relation of Cappellari et al. 2006); the color code represents the fraction of AGN as a function of $M_{BH}$ for any given value of $M_*$, as indicated by the color bar. We also show some of the paths in the $M_{BH}(t)-M_*(t)$ plane followed, during their evolution, by BHs (and by their host galaxies) reaching a final  mass of $M_{BH}(z=0)\geq 10^{9}\,M_{\odot}$.}
\vspace{0.4cm }

\subsection{The distributions of the Eddington ratio}

The different paths characterizing the BH growth in the IT and DI models suggests that the specific growth rate $\dot M_{BH}/M_{BH}$ could constitute an effective quantity to discriminate between the two modes; thus, we investigated how the distributions of Eddington ratio $\lambda=L_{AGN}/L_{Edd}$ (here $L_{Edd}\propto M_{BH}$ is the Eddington luminosity,  and $L_{AGN}\propto \dot M_{BH}$ after eq. \ref{Lagn}) compares with present observations. 
 The result is presented in fig. 7, where the AGN bolometric luminosity is 
shown as a function of the BH masses for the IT and DI scenarios, in two redshift bins.  The dashed lines correspond to Eddington ration $\lambda=1$, 
$\lambda=10^{-1}$, and $\lambda=10^{-2}$. Here the distributions appear substantially different when the IT or the DI modes are assumed. In the former case, a large spread is present for most of the explored luminosity range, while the average value of $\lambda$ increases with increasing $L_{AGN}$, to reach values $\lambda\approx1$ at the highest luminosities. Conversely, for the DI scenarios the distributions are characterized by average values of $\lambda$ increasing for decreasing luminosities, and by a small scatter.  This is not unexpected, due to the tighter region of the parameter space (gas rich $f_{gas}\sim 1$, disk dominated $f_d\sim 1$ dominated galaxies) where the DI mode is expected to be effective after eq. \ref{hopkins}, and also to the smaller variance associated to the merging histories when an internal trigger is assumed, as we discussed above.  We compare our predictions with data from Rosario et al. (2013), derived from spectroscopically observed AGN in the COSMOS field  at 0.5$<z<$2.2 with reliably black hole mass estimates from multiple optical spectroscopic surveys: the Sloan Digital Sky Survey (DR7) (Abazajian et al. 2009), the zCOSMOS bright and deep surveys (Lilly et al. 2007), Magellan/IMACS program  (Trump et al. 2009); BH masses were computed using virial relationships, and  H$\beta$ and MgII emission lines (Trakhtenbrot \& Netzer (2012), while 
bolometric luminosities  were estimated using bolometric corrections to the monochromatic luminosities at either 5100$\AA$  or 3000$\AA$  rest-frame. The choices of bolometric
corrections are derived in Trakhtenbrot \& Netzer (2012) and
are consistent with the prescriptions of Marconi et al. (2004). 

Note that at low AGN luminosities the comparison with data is affected by the observational bias set by the flux limit of the sample. This corresponds to L$_{AGN}\sim10^{44.2}$ erg/s at $z=0.5-1$ and  $L_{AGN}\sim 10^{44.7}$ erg/s at $z=1.5-2$. At low redshift $z=0.5-1$, where the flux limit could still allow to probe the low-luminosity end of the distribution, the comparison with the model is affected by the paucity of data points; this is due to a further selection bias, since the zCOSMOS AGN, which statistically dominate the of the sample, were specifically chosen to include the MgII line, which enters into the wavelength range of the zCOSMOS bright spectra at z$\sim$1.

\begin{center}
\vspace{0.2cm}
\scalebox{0.4}[0.4]{\rotatebox{0}{\includegraphics{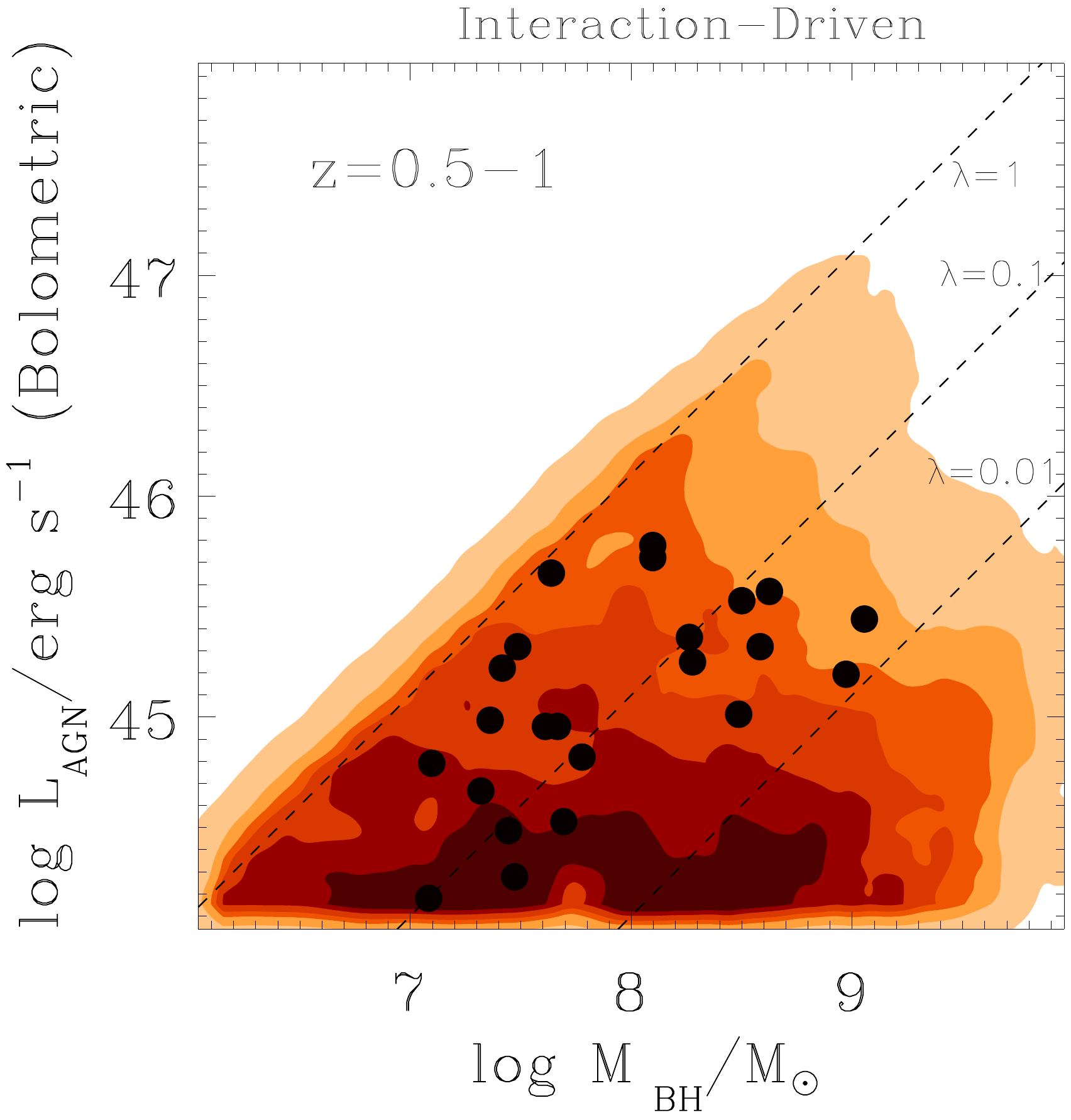}}}
\hspace{1cm}
\scalebox{0.4}[0.4]{\rotatebox{0}{\includegraphics{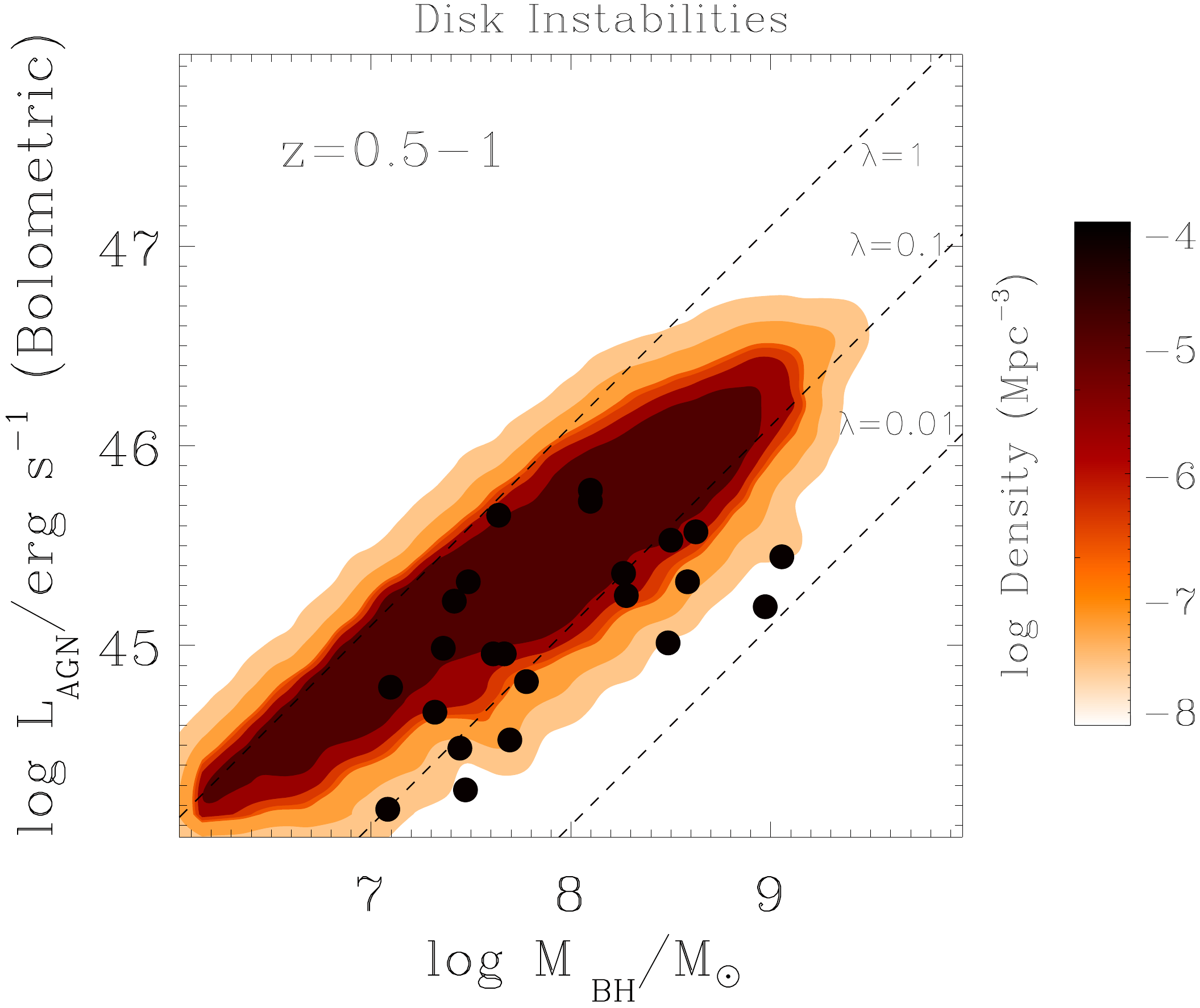}}}
\scalebox{0.4}[0.4]{\rotatebox{0}{\includegraphics{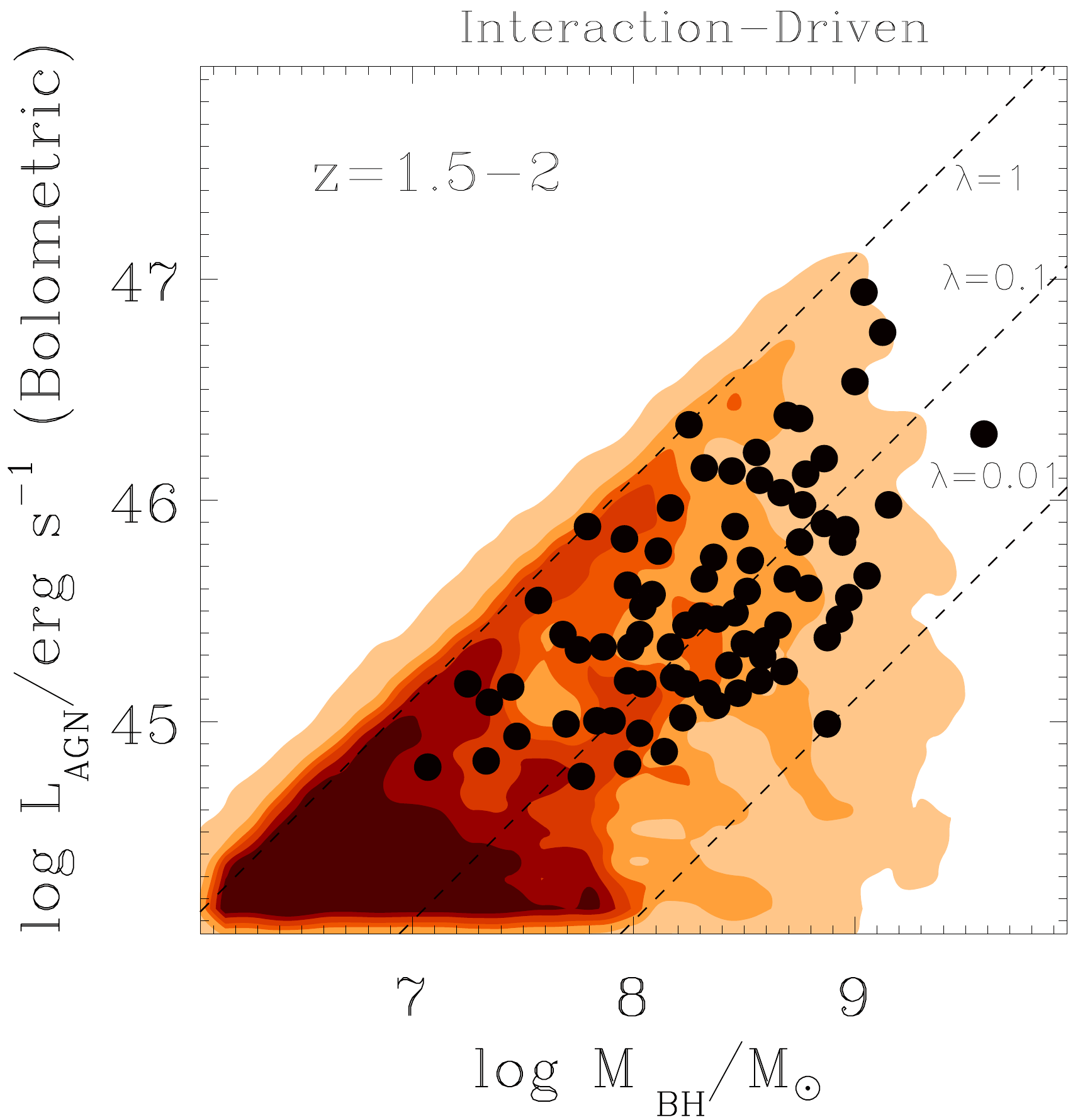}}}
\hspace{1cm}
\scalebox{0.4}[0.4]{\rotatebox{0}{\includegraphics{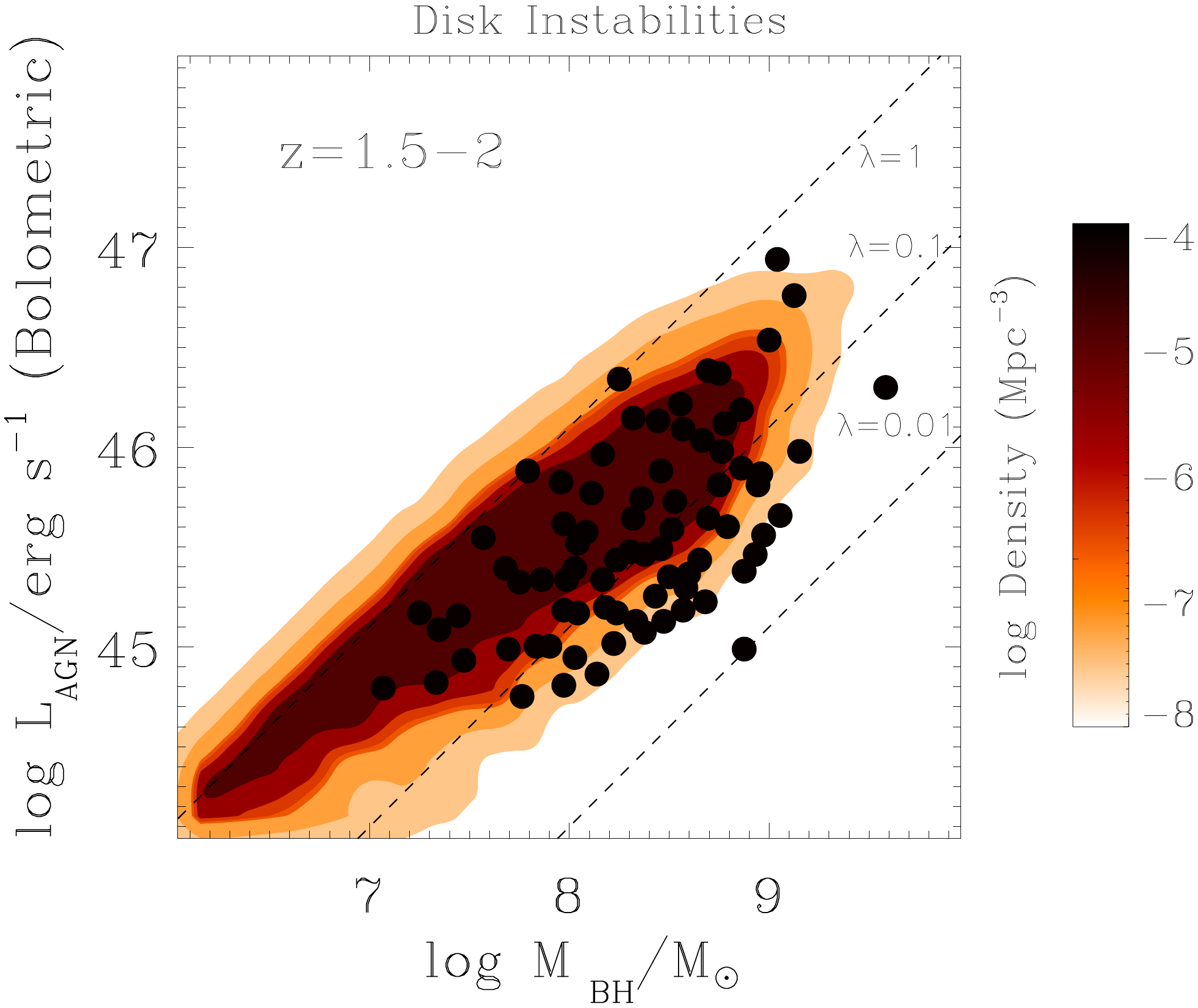}}}
\end{center}
\vspace{-0.2cm }
 {\footnotesize 
 Fig. 7. -The predicted distribution of AGN in the $L_{AGN}-M_{BH}$ plane for the IT  (left) and DI (right)  models is compared with data from Rosario et al. (2013) in two 
 redshift bins. The lines corresponding to Eddington ratio $\lambda=1$, $\lambda=0.1$ and $\lambda=0.01$ are also indicated. 
\vspace{0.3cm}}

Since the distribution of AGN in the $L_{AGN}-M_{BH}$ plane is appreciably different for the DI and IT case, a possible way of determining which of the two modes dominates the accretion for  intermediate- and low-luminosity AGN could consist in comparing the corresponding distributions of Eddington ratios $\lambda$ with present observations at low redshifts, where low-luminosity objects are accessible to observations; in fact, for such AGN, the IT mode is expected to yield distributions with larger variance and skewed to values $10^{-2}\lesssim \lambda\lesssim 10^{-1}$,  
while larger values of $\lambda$ and a smaller scatter are expected for the DI mode. 

However, when the model predictions for the two modes are compared with data  (see fig. 8) we face with a complex situation;  in fact, the observational distributions (at least) depend on i) the method adopted to select AGN; ii) the adopted bolometric correction; iii) the method adopted to measure the BH mass  (for a discussion of the above points, see, e.g., Netzer et al. 2009a,b and references therein). This is well illustrated by the  top-left panel, where we compare with low-redshift data  for AGN with $L_{AGN}<10^46$ erg s$^{-1}$ taken from Hickox et al. (2009) and Kollmeier et al. (2006). The former refer to AGN at 0.25 $< z <$ 0.8 selected in different wavebands: the radio data are taken from the Westerbork Synthesis Radio Telescope 1.4 GHz radio survey (de Vries et al. 2002),  the X-ray data are from the XBootes survey, covering the full AGES (AGN and Galaxy Evolution Survey) spectroscopic region (Murray et al. 2005; Kenter et al. 2005), while the Mid-Infared data are from the Spitzer IRAC Shallow
Survey covering the full AGES field in all four IRAC bands (Eisenhardt et al. 2004).
To estimate $M_{BH}$ the authors base on the $L_{B,bul}-M_{BH}$ relation after estimating the bulge luminosity of the host galaxy in the B band, while the bolometric luminosity for X-ray AGN are derived  from the bolometric correction of Hopkins et al. (2007); for radio AGNs that are not detected in X-rays, they used the X-ray stacking results  to derive approximate upper limits on the X-ray AGN luminosity, while for IR AGNs that are not detected in X-rays, L$_{bol}$  is derived from the rest-frame 4.5 $\mu$m luminosity. We also compare with data from Kollmeier et al. (2006), derived from the AGES survey and  consisting of X-ray (XBootes survey, Murray et al. 2005, Kenter et al. 2005)  or mid-infrared (Eisenhardt et al. 2004) point sources with optical magnitude R$\leq$21.5 mag and optical emission-line spectra characteristic of AGNs. Black hole masses  were estimated using virial relationships and H$\beta$ MgII, and CIV emission-line widths, while bolometric luminosities  were estimated using the bolometric correction of Kaspi et al. (2000) to the monochromatic luminosities  at 5100 $\AA$. 
 
\begin{center}
\vspace{-0.1cm}
\scalebox{0.6}[0.6]{\rotatebox{-90}{\includegraphics{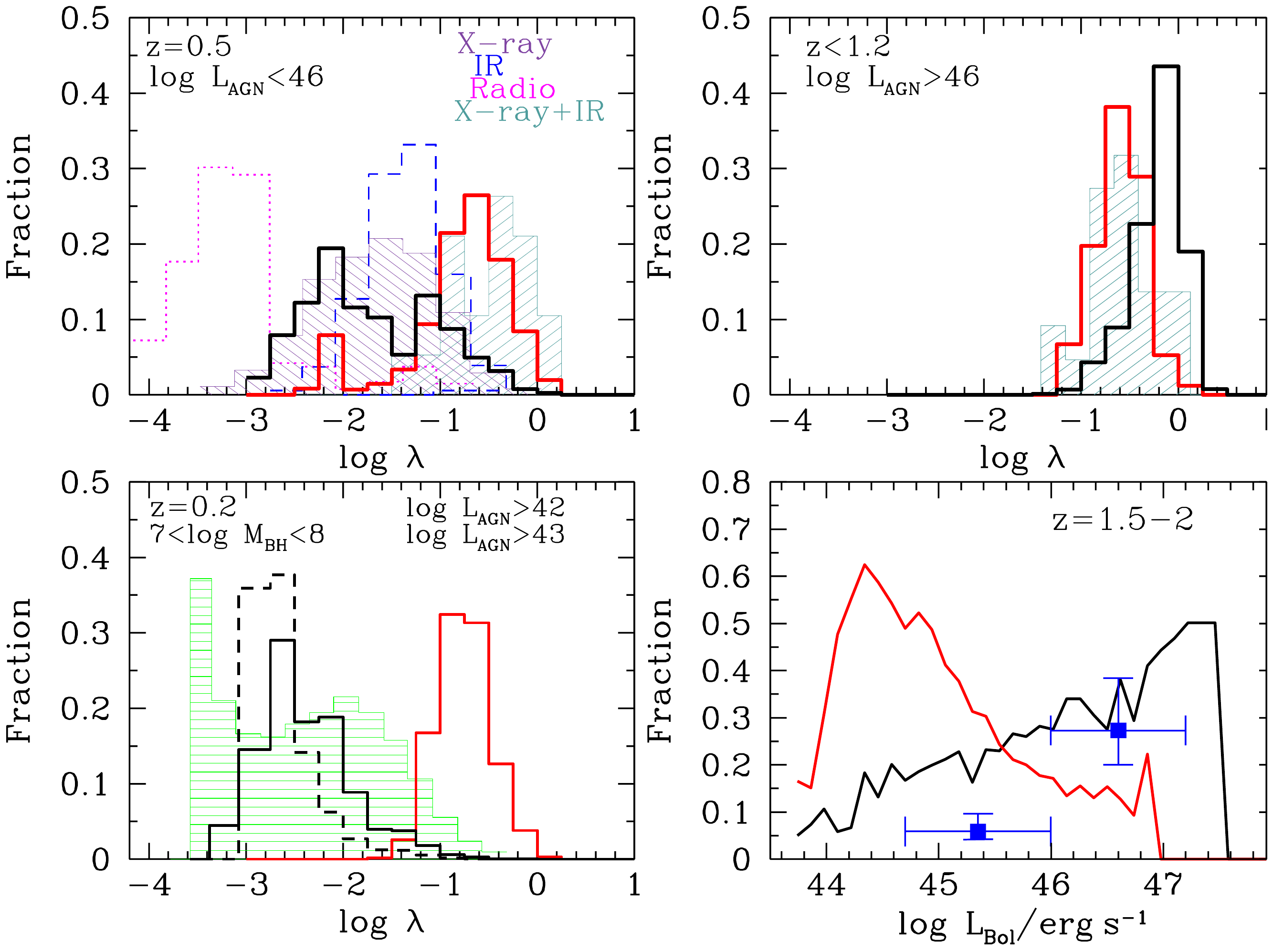}}}
\end{center}
\vspace{-0.cm }
 {\footnotesize 
Fig. 8. - Top Left Panel: The predicted distributions of the Eddington ratio $\lambda$ in the IT (solid black line) and DI (solid red line) scenarios for AGN with bolometric luminosity $L_{AGN}<10^{46}$ erg s$^{-1}$ at $z\leq 0.5$. They are compared with different sets of data (see text for details): 
a sample of X-ray selected AGN (dashed violet histogram), of IR selected (dashed histogram) and Radio 
selected AGN (dotted histogram) from Hickox et al. (2009), and the sub-sample of X-ray and IR AGN from Kollmeier et al (2006, dark green histogram) corresponding to $z\leq 0.5$. 
Top Right Panel: As above, but for luminous AGN with $L_{AGN}>10^{46}$ erg s$^{-1}$ at $z\leq 1.2$. 
The dashed histogram represents observational results from Kollmeier et al (2006). \\
Bottom Left Panel: As above, but for AGN with BH masses in the range $10^7\,\,M_{\odot}\leq M_{BH}\leq 0^8\,\,M_{\odot}$. Here we compare with data from Kauffmann \& Heckman (2009, histogram). 
Bottom Right Panel: The fraction of AGN with high Eddington ratios λ$\lambda\geq  0.5$ as a function of the AGN bolometric luminosity $L_{AGN}$, at $z=2$. As above, solid black line  corresponds to the IT mode  while solid red line to the  DI case. The points are computed from the data by Rosario et al. (2013) shown in fig. 7. 
\vspace{0.3cm}}

The comparison in fig. 8 (top -left panel) shows that, while a marked difference is indeed apparent between model predictions for the IT mode and the DI mode, the observational distributions for X-ray selected AGN (Hickox et al. 2009) extend to lower values of $\lambda$ with respect to IR AGN, while Radio selected objects show a distribution peaked to much lower values of $\lambda$. While the X-ray homogeneous sample in Hickox et al. (2009) would favor the IT accretion mode, the mixed IR+X-ray data by Kollmeier et al. (2006) are instead best matched my the DI mode predictions, showing the impact of adopting different methods to measure BH masses on the comparison with model predictions.  Note again how that scatter constitutes a crucial diagnostics to pin down the dominant accretion mode for low- and intermediate- luminosity AGN (see also Shankar et al. 2013).  Much weaker constraints come instead from luminous AGN (top-left panel), where the two modes yield similar distributions and scatter; for luminous AGN, however, we know from the results in fig. 4 that IT mode must dominate, since accretion from DI does not provide the observed abundance of bright AGN. 

We also compare (fig8, bottom-left panel) with data for low-mass BHs ($10^7\leq M_{BH}/M_{\odot}\leq 10^8$) taken from Kauffmann \& Heckman (2009); their  sample is drawn from the Data Release 4 of the SDSS at $z\simeq$0.2 (York et al. 2000; Adelman-McCarthy et al. 2006). AGN are identified from their position in the  [O III]($\lambda$=5007$\AA$)/H$\beta$ versus
[N II]($\lambda$=6583$\AA$)/H$\alpha$  diagram, as discussed in Kauffmann et al. (2003). Here $L_{AGN}$ is derived from the luminosity $L_{OIII}$ of the [OIII]($\lambda=$5007$\AA$) line after applying a bolometric correction that they estimate in the range 500-800, while BH masses were estimated using the stellar velocity dispersions measured from the SDSS spectra and the formula given in Tremaine et al. (2002). Within the limitations of available data, affected by a substantial uncertainties concerning the bolometric correction, the measurements of the BH mass, and the selection effects,  the spread and the peak of the observed distribution seem to favor the IT scenario, but a conclusive evidence is still missing. Note also that $\lambda$-distributions with a large spread is also indicated by analysis of the observed AGN duty cycles (see Shankar et al. 2013). 

Thus, at present the observational situation does not allow to draw a definite discrimination between the 
DI and the IT accretion modes. The key point is to assess whether low- and intermediate- luminosity AGN have a narrow distribution of $\lambda$  peaked at $\lambda\sim 1$ (as predicted for the DI mode) or a wider distribution extending to $\lambda\sim 10^{-1}-10^{-2}$. 
Thus, a straightforward observational test is constituted by 
 the fraction of AGN with high ($\lambda\geq 0.5$) Eddington ratios as a function of the AGN luminosity (choosing a redshift bin where significant statistics is ensured,  $2\leq z\leq 3$). An opposite behaviour is then predicted when IT and DI  are assumed as dominant accretion modes, as shown in the bottom-right panel of fig. 8. We stress  again that, while the dominance of the IT mode for bright AGN results from the luminosity functions in fig. 4, the interesting test is constituted by the low-luminosity 
 region of the plot, where an increasing fraction for lower luminosities would clearly point toward the effectiveness of DI in triggering AGN; however, present data (taken by integrating the $L_{AGN}-M_{BH}$ data by Rosario et al. (2013) shown in fig. 7) - although consistent with a dominance of the IT mode for luminous AGN - still do not allow to probe the low-luminosity range. 
Measuring such a quantity and comparing it with the predicted distributions could provide a clear observational test to assess whether DI may be the dominant accretion mode for intermediate-low AGN luminosities. 

\section{Discussion and Conclusions}
Using a a state-of-the-art semi analytic model for galaxy formation we have investigated in detail the effects of accretion triggered by disk instabilities 
(DI) in isolated galaxies on the evolution of the AGN population and on the corresponding growth of massive BHs. Specifically, we took on, developed and expanded the Hopkins \& Quataert (2011, HQ11) model for the mass inflow following disk perturbations, based on a physical description of nuclear inflows and 
 tested against aimed N-body simulations.

We have compared the evolution of AGNs due to such a DI accretion mode with that arising in a scenario where galaxy interactions produce the sudden destabilization of large quantities of gas feeding the AGN (interaction-triggered - IT - mode);  
this constitutes the standard AGN feeding mode implemented in the earliest versions of our semi-analytic model,  and has been extensively tested in our previous works. In the present work we do not attempt to develop a best-fitting model including both accretion modes, but rather we investigate the effects of assuming DI or IT as dominant modes studying the {\it maximal} contribution of DI to the evolution of the AGN population, within the framework of the HQ11 model for nuclear inflow. To this aim, we extended and developed the HQ11 model for DI, to assess the effects of changing the assumed disk surface density profile and the associated nuclear star formation rates within the limits provided by N-body simulations. The scope is to investigate how often and to what extent the galaxy properties evolving in a cosmological context (and described through  a semi-analytic model) provide the nuclear inflows associate to disk instabilities in ultra-high resolution simulations. We obtained the following results:

$\bullet$ For luminosity $M_{1450}\gtrsim -26$ AGN the DI mode can provide the BH accretion needed to match the observed AGN luminosity  functions up to $z\approx 4.5$; in such a luminosity range and redshift, it constitutes a viable candidate mechanism to fuel AGN, and can compete with the IT scenario as the main driver of cosmological evolution of the AGN population.

$\bullet$ The DI scenario cannot provide the observed abundance of high-luminosity QSO with $M_{1450}\leq -26$ AGN, as well as the abundance of 
high-redhshift $z\gtrsim 4.5$ QSO with $M_{1450}\leq -24$. This is because the strong dependence of the BH accretion rate on the gas fraction $f_{gas}$ and on the B/T ratio in the HQ11 model limits the effectiveness of such an accretion mode in massive galaxies; at high redshifts, the frequent major mergers predicted to take place at high redshifts $z\gtrsim 5$ in the host galaxies cause the rapid destruction of disks, and then lower the accretion rate in the DI mode due to its strong dependence on the disk mass fraction $f_d$. As found in our earliest works, the IT scenario provides an acceptable match to the observed luminosity functions up to $z\approx 6$. 

$\bullet$ The above conclusions concerning the DI mode do not change when a different (exponential) surface density profile is assumed for the disk with respect to the original HQ11 model, or when nuclear star formation rates higher than the lower limits derived in the framework of the HQ11 model are assumed. Indeed, the latter effect decreases the cold gas supply in the host galaxies and hence suppresses the effect of the DI accretion rates, leading to lower luminosity functions. 

$\bullet$ The BH mass distribution predicted in the DI and IT scenarios at $z\lesssim 0.3$ are consistent with present observational constraints, although the IT model provides a better fit to the high-mass end. Also, the observed local $M_{BH}-M_*$ relation is consistent with that predicted in both the DI and IT scenarios. However the DI scenario provides a slightly steeper slope of the relation, and - most of all - an appreciably smaller scatter. Such a difference is due to the larger variance in the paths followed by the AGN hosts to reach their final position in the $M_{BH}-M_*$ plane in the IT scenario; in fact, in such a case   the trigger and the size of the BH accretion directly depend on the merging histories, and are thus heavily affected by the associated variance. On the contrary, in the DI case the paths are confined in a small region of the plane; the growth of BH follows a corresponding growth of the stellar mass content, and phases where either $\dot M_{BH}$ or $\dot M_*$ largely dominate are absent. This means that we expect a tight correlation between the AGN luminosity and the star formation rate in the host galaxy. This constitutes an interesting point that we shall investigate in detail in a following paper.

$\bullet$ The distribution of the Eddington ratio $\lambda$ can constitute an effective probe to pin down the dominant fueling mechanism of AGN 
in the  low-intermediate luminosity range $L_{AGN}<10^{46}$ erg s$^{-1}$, where the DI and the IT modes are both viable candidates as the main drivers of the AGN evolution. 
In fact, for such AGN, the IT mode is expected to yield distributions with larger variance and skewed to values $10^{-2}\lesssim \lambda\lesssim 10^{-1}$,  
while larger values of $\lambda$ and a much smaller scatter are expected for the DI mode. 
Note that, tuning the DI mode as to yield lower values for the Eddington ratio $\lambda$ would result in a severe under-prediction of the abundance of AGN. The comparison of the model predictions  with observational $\lambda$-distributions at low redshifts, where the population of low-luminosity is accessible, is still plagued by the combination of selection effects and uncertainties concerning the bolometric corrections and the adopted method to measure the BH mass, so present data still do not allow to draw firm conclusions. To alleviate the effects of the above biases and uncertainties by 
increasing the statistics, we propose a test  at $z\gtrsim2$ based on the fraction of low-luminosity AGN ($L_{AGN}\leq 10^{44}$ erg s$^{-1}$) with high Eddington ratios $\lambda\geq 0.5$ at $z\gtrsim2$.  The presence  of a peak at low luminosities ($L_{AGN}\lesssim 10^{44}$ erg s$^{-1}$) would indicate a  significant contribution of the DI mode to the growth of supermassive BHs.
Finally, we note that the our results on the small scatter of the $\lambda$-distribution in the DI scenario compared to the IT mode agree with those obtained from semi-empirical studies by Shankar et al. (2012; 2013). 

We stress the DI and IT modes actually constitute schematic representations of the actual AGN fueling process. The first case applies to isolated galaxies, where  perturbations $a\approx 0.1-0.3$ of the potential $\Phi_a=a\,\Phi_0$ with respect to the axisymmetric thin disk potential $\Phi_0$ develop 
when the gas mass exceeds a critical value (eq. 4); in this case, the HQ11 model describes the corresponding mass inflow attainable in the galaxy nucleus. 
On the other hand, the IT model corresponds to a scenario where the large quantities of gas destabilized during encounters cannot be regarded as disk perturbations, and derives the corresponding nuclear feeding from the model by Cavaliere \& Vittorini (2000). Of course, intermediate situations may arise in minor merging events, where the gas destabilized during encounters or fly-by can be regarded as a perturbation of the potential, and hence it is subject to the accretion limits give by the HQ11 model; this constitutes an interesting development of the model that we shall explore in a next paper. 

To what extent the HQ11 description implemented in our semi-analytic model represent the observed features generally associated to disk instabilities ? As for the star formation properties of DI hosts, Silverman et al. (2009) studied the correlation between $\dot M_{BH}$ and the total star formation in galaxies hosting a AGN in isolated (non-merging) systems. Using a large sample of galaxies from the zCOSMOS survey they select a sample of moderately luminous AGN with no structural signs of recent interactions with other galaxies, the best candidates to probe the  DI scenario. They find that median ratio $\dot M_* /\dot M_{BH}\approx 10^2$  almost constant in the redshift range $z \sim 0.5 - 1 $; such a ratio is in good agreement with $A_{*,DI} = 100$ of our fiducial DI case computed in Appendix on the basis of the HQ11 model, and with that resulting from the simulations by Hopkins \& Quataert (2010) and Bournaud et al. (2011).  As for the morphological properties of AGN in the DI scenario, Bournaud et al. (2012) compare a sample of 14 ′′clumpy′′ galaxies at z=0.7 with a sample of 13 stable and smoother galaxies
at the same redshift; since those clumpy galaxies are found to be on the Main Sequence of star formation and do not show clear signs of recent interactions, the authors infer that DI rather than mergers may be responsible for such activity. Using emission line diagnostic, they find that clumpy disks have a higher probability of containing moderately luminous AGN (bolometric luminosity between $10^{43}$ erg s$^{-1}$ and $10^{44}$ erg s$^{-1}$. Since the most powerful accretion episodes driven by DI are thought to be associated with clumpy morphologies and high gas fractions (see, e.g., Ceverino, Bournaud, Dekel  2010; Bournaud et al. 2011), the observed morphologies of low-intermediate AGN hosts are consistent with those expected in the DI description implemented here (sect. 3.2.1). Thus, the description of DI adopted here and implemented in our semi-analytical model seems to capture the morphological and star formation signatures generally associated to AGN powered by DI.  

How robust are our conclusions concerning the inefficiency of DIs in powering bright QSOs ? The drop of the effectiveness of DI in accounting for bright AGNs is due to the strong dependence on the B/T ratio (or, equivalently, on $f_d$)  in eqs. 7 and 8. 
Such a dependence has been measured in simulations under a wide range of initial conditions and of galaxy properties (such as equation of state of the inter-stellar medium, 
and Supernovae feedback, see Hopkins \& Quataert 2010), so it seems to constitute a robust feature of DI accretion mode. In addition, 
we adopted values $a=0.3$ for the amplitude of disk disturbances in the DI case; this actually constitutes an upper limit to values measured in simulations which actually range from $10^{-2}$ to $3\,10^{-1}$ (Hopkins \& Quataert 2010), so again the effects of DI on the luminosity functions in fig. 4 are actually upper limits. 
Indeed, Bournaud et al. (2011) study disk galaxies at $z\approx 2$ using both analytical and numerical estimates for the accretion rate onto the central BH in galaxies with high gas surface density continuously fed by gas streams. In their aimed and isolated galaxy simulations, they start from highly unstable  ($Q\, < \,1$) and gas rich disks ($f_{gas} \sim 0.5 $);  the gas reservoir is supposed to be continuously fed by cold gas streams and gravitational instability play a crucial role for feeding the central BH. Assuming $M_{BH}/M_{blg}  \sim 3 × 10^{-3}$ at z $\sim$ 2, for a galaxy of barionic mass $10^{11} M_{\odot}$ and $M_{BH} \sim 10^8 M_{\odot}$ they obtain from their analytic estimate $\dot M_{BH} \sim 0.04 M_{\odot}yr^{-1}$ and a corresponding bolometric luminosity  $L_{AGN}\approx 2 \, 10^{44}$ erg s$^{-1}$, corresponding to modest X-ray luminosities $L_X\approx 10^{42}-10^{43}$ erg s$^{-1}$. This is also the range of luminosities where the observational support of the relevance of DI in isolated galaxies is stronger, as noted above (see, e.g., Silverman et al. 2009, Bournaud et al. 2011). It is interesting to note that simulations suggest the accretion episodes triggered by DIs is spread over a longer period (up to 10 times) compared to  IT events  (Bournaud et al. 2011); this constitutes an interesting issue that we plan to take on in a next paper. 

We have also investigated the robustness of our results to changes in the modelling of  the bulge growth during the nuclear starbursts associated to the gas inflows. Since all the inflowing gas is either feeding the BH or converted into stars, we only need to verify the effects of varying the assignment of such stars to the bulge or to the disk component. Performing such a check yields no appreciable variation in  our results, even in the DI case.  In fact, although the bulge indeed stabilizes the disk (eqs. 8 and 9 show that increasing B/T yields lower inflows 
due to disk instability),  the contribution of starbursts to the stellar budget of the model galaxies is minor 
in the IT case (as shown by Lamastra et al. 2013a) and even more 
in the DI model, as also supported by observational data (see Rodighiero et al. 2011), except at very high 
redshifts $z>4$ where the IT mode (not  depending on the partition of stars in bulge or disk) is dominant anyway. 

Note that the above considerations imply that the growth of the bulge cannot effectively act as a 
suppression factor to provide low-luminosity AGN in galaxies with massive BHs, which are not produced 
in the DI scenario (see fig. 7).  In our cosmological framework, such objects can be naturally yielded in the IT scenario, since massive BHs  reside (on average) in large-mass galaxies which a) at low redshift accrete mainly small clumps, and b) have formed from condensations collapsed in biased, high-density regions of the density field and thus have converted most of their gas into stars (and in the BH feeding) at high redshift, leaving little fuel for the subsequent star formation and BH feeding processes. The triggering interactions correspond to minor merging (smooth accretion of small lumps, with masses as small as $10^{-2}$-$10^{-3}$ the host mass) and fly-by events. It is not easy to observationally  catch one of such events in the act since they  produce intrinsically less disturbance to the host galaxy making them hard to identify morphologically, and the systems can appear disturbed only for a short period of time (compared with the AGN active phase, see Storchi-Bergmann et al. 2001). In fact, they are not expected to yield an appreciable enahncement in the clustering of AGN hosts (see Taniguchi 1999; see also Stewart 2009; Lotz et al. 2011).  Thus the controversial observational results concerning the clustering of the AGN hosts - see  Li et al. (2006),  Pierce et al. (2007), Darg et al. (2010) for a negative evidence, and Alonso et al. (2007), Woods \& Geller (2007),  Rogers et al. (2009), Koss (2010), and Ellison (2011) for an enhanced fraction of close pairs of AGN hosts compared to control samples - can hardly provide a definite clue for the minor merging/fly-by scenario  (also because of the dependence of the results on the clustering scale). A support for  the interaction-based  picture can be found in the bluer colors of galaxies close to AGN hosts (see Shirasaki et al. 2011; Kollatschny, Reichstein,  Zetzl 2012), and in the observations of specific individual cases (see, e.g., González Delgado et al. 2002; Anderson et al. 2013 and references therein). Finally, we note that 
an alternative (or complementary) possibility is the direct accretion of hot halo gas discussed and implemented in Croton et al. (2006). 

As for the relation between inflow and star formation, we recall that an equilibrium among the two is assumed in  HQ11 model to derive the limit accretion rates attainable in the nucleus. What would happen if such equilibrium were not respected ? Actually, different scenarios are possible. First of all, consider the case of a huge mass inflow into the central region of the galaxy: if the gas flows rapidly, it is possible that star formation is not able to contrast the inflow and equilibrium is not reached. Since the  surface density of star formation rate $\dot \Sigma_*$ decreases with increasing radius (eq. \ref{sigmastarin2}, \ref{sigmastarin3} and \ref{sigmastarout2}), we expect that only the very inner region ($R < R_{BH}$) could suffer from this kind of problem. Furthermore, since the star formation rate is no more able to contrast the inflow, we can assume that $\dot \Sigma_*$ could reach a maximum and constant value. Thus, the mass inflow can be simply evaluated using eq. \ref{massinflow} with $\Sigma_{g} \simeq const$ instead of using the equilibrium equation (at least in the very inner region). If we use eq. \ref{massinflow} to evaluate the mass inflow,  we will obtain a mass inflow $ \dot M \propto R^{\frac{3}{2}}$, actually steeper than that obtained when equilibrium is reached ($ \dot M\propto R^{\frac{5}{6}}$). Since out of a certain radius the mass inflow is the same for both cases, moving towards the centre the inflow will drop more if the $\dot M_*$ saturates leading to a less effectiveness of the final inflow rate onto the central BH. Thus, assuming that equilibrium is always reached provides {\it upper limits} to the  real mass inflow rate, so our basic conclusions remain valid. 
On the other hand, it is also possible that when gas density is sufficiently low, both mass inflow and/or star formation could be altered. But since low $\Sigma_{g}$ is in general correlated with low mass inflow (eq \ref{massinflow}), in these cases star formation is usually more effective and consumes most of the gas, suppressing the mass inflow; again, considering the equilibrium, we are actually overestimating the mass inflow rate.

A final fundamental question that remains open concerns the balance between the IT and the DI modes 
for low-intermediate luminosity AGN. In fact, in such a luminosity range, fig. 4 shows that both modes could provide the bulk of the AGN population. Although disentangling the contributions from the two is not an easy task, the results in the present paper open the way to several interesting possibilities. One specific prediction that will allow to constrain the contribution DIs to the BH growth is constituted by the luminosity distribution of hosts with high Eddington ratio (see fig. 8). Other predictions concern the small variance in the paths followed by galaxies in evolving in the $M_{BH}-M_*$ plane if BH are fed by DIs; this means that in such a scenario BH growth $\dot M_{BH}$ and 
star formation $\dot M_*$ are tightly related, a situation which can be readily tested by comparing the model predictions with observed 
$L_{AGN}-\dot M_*$ relations, or with the observed color distribution of AGN hosts. We plan to develop in detail such comparisons in a next paper. 

\section*{Acknowledgements} This work was supported by ASI/INAF contracts I/024/05/0 and I/009/10/0 and PRIN INAF 2011. 

\newpage
\appendix
\section{BLACK HOLE ACCRETION RATE AND NUCLEAR STAR FORMATION RATE FROM DISK INSTABILITIES: THE CASE OF POWER-LAW SURFACE DENSITY PROFILES}
Following the computation in HQ11, we solve eqs   \ref{massinflow} and \ref{masstot} for both $\Sigma_g$ and $\dot M_{infl}$. To this  
aim, we consider two distinct regions: an outer region where the  potential is dominated by the disk, and an inner region where the BH gravity is dominant. The radius $R_{BH}\sim $ 10 pc marks the separations between the two regions, while the scale $R_{acc}\sim 0.1$ pc corresponds to the inner bound of the innermost regions where we want to compute the BH accretion rate in eq. 
\ref{hopkins}. 

\subsection{\bf The Potential}

 To estimate the potential in the outer regions (from $\sim$ 10 pc to $\sim$ 100 pc),  HQ11 adopt the WKB approximation to estimate the perturbation amplitude $\Phi_a \approx |a|_{max} 2 \pi G \Sigma_d R |kR|^{-1}$, taking $|kR| \sim 1$ since they are considering global modes; on the basis of aimed N-body simulations (Hopkins \& Quataert 2010a)  the dominant mode is assumed to be $m=2$ and the maximal perturbation amplitude $|a|_{max} \approx 0.3 f_d(R)$ is taken  to be proportional to the disk fraction $f_d(R)$
for any radius $R$. In the inner one (from $R_{acc}$ $\sim$ 0.1 pc to $R_{BH}$ $\sim$ 10 pc) the potential is dominated by the BH and $\Phi_a \approx |a|_{max} G M_d(R_{BH})/R_{BH}$, with $|a|_{max} \approx 0.2 f_d(R_{BH})$. In this region, the dominant mode is $m=1$, also motivated  by simulations. 

\subsection{\bf The Gas Surface Density}

In both regions, disk, bulge and gas surface densities are assumed to be described by power-law profiles, that is $\Sigma_{d} \propto R^{-\eta_d}$, $\Sigma_{b} \propto R^{-\eta_d}$, $\Sigma_{g} \propto R^{-\eta_g}$. In addition, HQ11 assume the disk to be stellar dominated so that the total 
surface density $\Sigma_t =\Sigma_b+\Sigma_d\propto  R^{-\eta_d}$, and the disk fraction 
$f_d=\int \Sigma_d\,R\,dR/\int \Sigma_t\,R\,dR$ is constant with radius.

With this assumptions it is possible to solve eq. \ref{massinflow} and \ref{masstot} for both the gas density $\Sigma_g$ and the 
mass inflow rate $dM_{infl}/dt$.  In particular, in the inner region $R_{acc}\leq R\leq R_{BH}$ where the BH dominates the potential $\Phi$ in eq. \ref{massinflow}, eq. \ref{masstot} yields:

\begin{align}
\label{sigmain}
\left( \dfrac{\Sigma_g}{\Sigma_k} \right)^{\eta_k-1}_{inner} = \dfrac{3 \eta_k - 4}{20 \pi (\eta_k-1)} \sqrt{\dfrac{GM_{BH}}{R_{BH}^3}}t_k \left(\dfrac{R}{R_{BH}} \right)^{-1/2}f_d &
\hspace{6cm}  & 
R_{acc}\leq R\leq R_{BH}
\end{align}

while in the outer region $R_{BH}\leq R\leq R_{0}$, where the potential $\Phi$ is dominated by the disk, the same equations yield:
\begin{align}
\label{sigmaoutfiducial}
\left( \dfrac{\Sigma_g}{\Sigma_k} \right)^{\eta_k-1}_{outer}= \alpha_d \left( \frac{\Sigma_d}{\Sigma_t}\right)^2 \Omega\,t_k ~~~~~~~~~~~{\rm     with}~~\alpha_d = \frac{m a_0 \,-2\eta_d -\eta_g+ 3(\eta_d+1)/2|\,(2-\eta_d)^2}{8 \pi (2 - (\eta_d+1)/2)} &  \hspace{1.5cm}  & R_{BH}\leq R\leq R_{0}
\end{align}
where $\Omega=\Big[G\,2\pi\,\int_0^R (\Sigma_d+\Sigma_b)\,R\,dR/R^3\Big]^{1/2}$ is the angular velocity of the disk. 

\subsection{\bf The Boundary Conditions}
To compute the accretion flow in the inner region $R_{acc}\leq R\leq R_{BH}$ we can  use 
eq. \ref{sigmain} to solve eq. \ref{masstot}. However, the two density profiles in eqs. \ref{sigmain} and \ref{sigmaoutfiducial} are not matched as to  ensure a continuous mass flux 
at the boundaries $R_{BH}$ and $R_0$. In principle, one could impose a continuity condition 
for such equations at all times; since the boundaries move as accretion proceeds, this would require a detailed treatment following the dynamical evolution of the boundaries and of the mass  
flux across them. Following HQ11, we adopt a simplified approach; we keep the density profiles 
in eqs. \ref{sigmain} and \ref{sigmaoutfiducial}, and we impose the mass conservation 
directly in eqs. \ref{masstot}. Thus, we write the latter as 
\begin{equation}
\label{masstot_bound}
{dM_{infl}\over dt}=2\,\pi\,R\,dR{\Sigma_k\over t_k}\,\Bigg({\Sigma_g\over \Sigma_k}\Bigg)^{\eta_k}\,A(R_{BH},R_0)
\end{equation}
where $A(R_{BH},R_0)$ expresses the proper boundary conditions, i.e., the mass 
conservation across the boundaries $R_{BH}$ and $R_0$. To compute such a quantity, we note that the mass accreted into the inner region ($R_{acc}\leq R\leq R_{BH}$) must equal the disk mass that has flown across $R_{BH}$. We indicate with $\Delta M_g$ the mass inside the boundary $R_{BH}$ 
computed after integrating eq. \ref{sigmain}, and with $A(R_{BH},R_0)\Delta M_g$ the 
the actual mass accreted in the inner region after matching with it with the mass flown from the disk. In a time interval $\Delta t$,  we must have 
\begin{equation}
{A(R_{BH},R_0)\Delta M_g\over \Delta M_g}=A(R_{BH})=
{\dot M_g(R_{BH},outside)\over \dot M_g(R_{BH}, inside)+
\int_{R_{acc}}^{R_{BH}}\,\Sigma_{g,inner}(R)\,2\pi\,R\,dR/\Delta t}~.
\end{equation}
We must have $\dot M_g(R_{BH},outside)=\dot M_g(R_{BH}, inside)$. Estimating exactly such a quantity would require a dynamical description of the variation of the boundary with time, and of the the mass flux across it. However, a simple - though approximate - estimate of such a 
quantity can be obtained computing it as the mass lost from disk during $\Delta t$ (divided by $\Delta t$); this can be obtained after integrating the 
disk density profile \ref{sigmaoutfiducial} extrapolated down to $R=0$, so that 
$\dot  M_g(R_{BH},outside)\approx B(R_0)\int_0^{R_{BH}}\Sigma_{g,outer}(R)\,2\pi\,R\,dR/\Delta t$.  The quantity $B(R_0)$ is the normalization corresponding to the boundary condition at the galactic scale $R_0$; it ensures that the mass flux outside the boundary $R_0$ matches with that in the  region $R_{BH}\leq R\leq R_0$.  We then obtain
\begin{equation}
\label{boundary1}
A(R_{BH},R_0)=
{ 1\over 1+\int_{R_{acc}}^{R_{BH}}\,\Sigma_{g,inner}(R)\,2\pi\,R\,dR/B(R_0)\int_{R_{acc}}^{R_{BH}}\,\Sigma_{g,outer}(R)\,2\pi\,R\,dR}={1\over 1+2.5\,f_d^{-4/3}/B(R_0)}
\end{equation}
where the last equality follows after performing the integrals over $\Sigma_{g,inner}$ and 
$\Sigma_{g,outer}$ given in eqs. \ref{sigmain} and \ref{sigmaoutfiducial}. Note that the quantity 
$A(R_{BH},R_0)$ acts like a suppression factor; when the mass inflow obtained using the (unmatched) 
density distributions \ref{sigmain} and \ref{sigmaoutfiducial} yields an inflow rate that {\it exceeds} the gas mass actually available from the disk at the position $R_{BH}$, it suppresses the mass flux as to maintain it lower or equal to such mass. In fact, for large values of 
$\Delta M_g(outside,R_{BH})$ we get $A(R_{BH},R_0)=1$ (no suppression is needed since 
the disk mass available from the disk is large). 

The quantity $B(R_0)$ in eq.\ref{masstot_bound} can be obtained iterating exactly the same mass-matching procedure, but applied to the outer boundary $R_0$; in terms of the disk mass $M_d(R_0)$ within $R_0$ and of the corresponding gas mass $M_g(R_0)=f_{gas}\,M_d(R_0)$ the result reads 
\begin{equation}
\label{boundary2}
B(R_{0})={ 1\over 1+\int_{R_{acc}}^{R_0}\,\Sigma_{g,outer}(R)\,2\pi\,R\,dR/
M_g(R_0)}=
{1\over 1+f_0/f_{gas}}~~~~~~{\rm with}~~~~~~~
f_0 \approx 0.2 f_d^2 \left( \dfrac{M_d(R_0)}{10^9 M_{\bigodot}}\right)^{-1/3}~~~~~
f_{gas} \equiv {M_g (R_0)\over M_d(R_0)}
\end{equation}

The physical meaning of the boundary factor $B(R_0)$ is to match the gas available in the galactic disk (given by the semi-analytic model) with that corresponding to the gas profile $\Sigma_g$ inside $R_0$. Note that, in the original HQ11 model, the gas density profile $\Sigma_g$ in the external region has a slightly higher normalization to interpolate between the marginal and the strong orbit crossing regimes; however, when such a factor is included, its effects on eq. \ref{boundary2} and eq. \ref{boundary1} balance, leaving our results unchanged. 

\subsection{\bf The BH Inflow Rate }

We now proceed to solve eq. \ref{masstot_bound} to derive the mass inflow rate $M_{infl}$ close to the inner bound $R_{acc}$.  HQ11 first consider eq. \ref{masstot} in an ring close to $R_{acc}$ and small enough that the surface density $\Sigma_g$ can be considered to be constant; this yields 

\begin{equation}
\label{masstot2}
\frac{dM_{infl}}{dt} (R_{acc})\simeq \pi R_{acc}^2 {\Sigma_k\over t_k}\,\Bigg({\Sigma_g(R_{acc})\over \Sigma_k}\Bigg)^{\eta_k}\,A(R_{BH},R_0)
\end{equation}

After substituting eq. \ref{sigmain} computed ad the inner radius $R_{acc}$ the above equation reads

\begin{equation}
\label{intern_inflow}
\frac{dM_{BH}}{dt} \equiv \frac{dM_{infl}}{dt} (R_{acc})= \pi \Sigma_k t_k^{\frac{1}{\eta_k-1}} \left(\dfrac{R_{acc}}{R_{BH}} \right)^{{3\eta_k-4 \over 2(\eta_k-1)}}  R_{BH}^2 \left[\dfrac{3 \eta_k - 4}{20 \pi (\eta_k-1)} \left(\frac{G M_{BH}}{R_{BH}^3}\right)^{\frac{1}{2}}f_d\right]^{\frac{\eta_k}{\eta_k-1}}\,A(R_{BH},R_0)
\end{equation}

Eq. \ref{intern_inflow} represents the mass inflow in the inner region corresponding to the BH accretion rate. We can relate $R_{BH}$ to $M_{BH}$ in terms of the disk mass within $R_{BH}$ through the relation $M_t(R_{BH}) \simeq M_{BH}$. The mass $M_t$ is computed as 
 $M_t(R_{BH})  = \int_0^{R_{BH}} 2 \pi \Sigma_tRdR$; for a power-law profile $\Sigma_t =A_{\Sigma_t} R^{-\eta_d}$ we can readily express   $R_{BH}$ as a function of $M_t=M_{BH}$ to obtain
\begin{equation}
\label{rbh}
R_{BH} = \left(M_{BH} \dfrac{2 - \eta_d}{2 \pi A_{\Sigma_t}}\right)^\frac{1}{2-\eta_d}
\end{equation} 
Substituting in eq. \ref{intern_inflow} yields 
\begin{equation}
\label{intern_inflow1}
\frac{dM_{BH}}{dt} = \pi \Sigma_k t_k^{\frac{1}{\eta_k-1}} 
\Bigg[\dfrac{3 \eta_k - 4}{20 \pi (\eta_k-1)}\, G ^{\frac{1}{2}}  \Bigg]^{\frac{\eta_k}{\eta_k-1}}
\left(\frac{R_{acc}}{R_{BH}} \right)^{3\eta_k-4 \over 2(\eta_k-1)} 
\,M_{BH}^{3\eta_k-4-\eta_k\eta_d\over 2\,(\eta_k-1)\,(2-\eta_d)}
f_d^{\frac{\eta_k}{\eta_k-1}}\,\left(\dfrac{2 - \eta_d}{2 \pi A_{\Sigma_t}}\right)^{\frac{\eta_k-4}{2(\eta_k-1)(2-\eta_d)}}\,A(R_{BH},R_0)
\end{equation}

We are now interested in expressing eq. \ref{intern_inflow} using physical quantities evaluated at far from the very central region, at "galactic scales $R=R_0\sim10^2-10^3$ pc, in order to derive the BH accretion rate from quantities computed from our semi-analytic 
galaxy formation model. To this aim, we note that, for any disk profile $\Sigma_d$, it is possible to combine $R_0$ and the associated enclosed disk mass $M_d(R_0)=\int_0^{R_0} 2 \pi \Sigma_d\,RdR$ to yield convenient constant quantities; in particular for a power-law profile, after some algebra, the last two factors in eq. \ref{intern_inflow1} can be recast as 
\begin{equation}
\label{const}
f_d^{\frac{\eta_k}{\eta_k-1}} \left(\dfrac{2 - \eta_d}{2 \pi A_{\Sigma_t}}\right)^{\frac{\eta_k-4}{2(\eta_k-1)(2-\eta_d)}} = M_d(R_0)^{\frac{4-\eta_k}{2(\eta_k-1)(2-\eta_d)}} R_0^{\frac{\eta_k-4}{2(\eta_k-1)}} f_d^{\frac{5\eta_k-4-2\eta_d\eta_k}{2(\eta_k-1)(2-\eta_d)}} 
\end{equation}
After substituting in eq. \ref{intern_inflow1} reads
\begin{equation}
\label{intern_inflow2}
\frac{dM_{BH}}{dt} = \pi \Sigma_k t_k^{\frac{1}{\eta_k-1}} 
  \left[\dfrac{3 \eta_k - 4}{20 \pi (\eta_k-1)}\, G ^{\frac{1}{2}}  \right]^{\frac{\eta_k}{\eta_k-1}}
\left(\frac{R_{acc}}{R_{BH}} \right)^{3\eta_k-4 \over 2(\eta_k-1)} 
\,M_{BH}^{3\eta_k-4-\eta_k\eta_d\over 2\,(\eta_k-1)\,(2-\eta_d)}
 M_d(R_0)^{\frac{4-\eta_k}{2(\eta_k-1)(2-\eta_d)}} R_0^{\frac{\eta_k-4}{2(\eta_k-1)}} f_d^{\frac{5\eta_k-4-2\eta_d\eta_k}{2(\eta_k-1)(2-\eta_d)}} \,A(R_{BH},R_0)
\end{equation}
We can express the above equation in physical units; this yields

\begin{equation}
\label{intern_inflow3}
\frac{dM_{BH}}{dt} = \alpha(\eta_k,\eta_d) \left(\frac{R_{acc}}{10^{-2}R_{BH}}\right)^{3\eta_k-4 \over 2(\eta_k-1)} 
\,\left(\frac{M_{BH}}{10^8 M_{\bigodot}}\right)^{3\eta_k-4-\eta_k\eta_d\over 2\,(\eta_k-1)\,(2-\eta_d)}
 \left(\frac{M_d(R_0)}{10^9 M_{\bigodot}}\right)^{\frac{4-\eta_k}{2(\eta_k-1)(2-\eta_d)}} \left(\frac{R_0}{100\,pc}\right)^{\frac{\eta_k-4}{2(\eta_k-1)}} f_d^{\frac{5\eta_k-4-2\eta_d\eta_k}{2(\eta_k-1)(2-\eta_d)}}\,A(R_{BH},R_0) { M_{\bigodot}\over yr}.
\end{equation}
where we have gathered all constant terms into the normalization 
\begin{equation}
\label{alpha}
\alpha(\eta_k,\eta_d)= \pi \left(0.4\right)^{\frac{1}{\eta_k-1}}\left(2.2 \,\dfrac{3 \eta_k - 4}{20 \pi (\eta_k-1)}\right)^{\frac{\eta_k}{\eta_k-1}}10^{\frac{32-15\,\eta_k -14\,\eta_d+7\,\eta_k\,\eta_d}{2(\eta_k-1)(2-\eta_d)}}
\end{equation}
The above expression is the BH accretion rate for generic power-law surface density profiles. 
After assuming a power-low index $\eta_d=1/2$ for the gas surface density profile, and an 
index $\eta_k=7/4$ for the Kennicut-Schmidt star formation law , the above equation yields eq. (9) in the main text and a normalization $\alpha(\eta_k,\eta_d)=0.18$. As noted by HQ11, assuming different values for $\eta_d$ and $\eta_k$ has only a minor effects on the exponents in eq. \ref{intern_inflow3}, but has a significant impact on $\alpha(\eta_k,\eta_d)$. For example, changing the $\eta_k$ to the value $\eta_k=3/2$ yields $\alpha(\eta_k,\eta_d)=3.17$. In the main text, we shall assume a reference value for $\alpha(\eta_k,\eta_d)=10$ to investigate the  {\it maximal} contribution of the DI mode to the evolution of the AGN population. 

\subsection{\bf The Nuclear Star Formation Rate }

The contribution to the star formation rate  can be calculated integrating star formation surface density rate $\dot\Sigma_*$ over the region of interest. The latter is 
related to the gas surface density by the Schmidt-Kennicut law $\Sigma_*\propto \left( \Sigma_g \right)^{\eta_k}$. For the inner region, $ \Sigma_g $ is described by eq. \ref{sigmain}; since all the terms in eq. \ref{sigmain} are constant but the radius, we obtain

\begin{equation}
\label{sigmastarin2}
\frac{d\Sigma_*}{dt} \propto R^{-\frac{\eta_k}{2(\eta_k-1)}}~.
\end{equation}

The normalization can be estimated considering the value that star formation rate  assumes in $R=R_{acc}$. This can be done easily since the mass inflow in $R=R_{acc}$ is strictly related to star formation surface density rate by eq. \ref{masstot} and gives $\dot M_{BH}= \dot M_{infl} (R_{acc})  \simeq \pi R_{acc}^2 \dot \Sigma_*(R_{acc})$. Thus, star formation surface density rate in the inner region is equal to
\begin{equation}
\label{sigmastarin}
\dfrac{d \Sigma_*}{dt} = \dfrac{dM_{BH}}{dt} \dfrac{1}{\pi R_{acc}^2} \left(\dfrac{R}{R_{acc}}\right)^{-\frac{\eta_k}{2(\eta_k-1)}}~.
\end{equation}
The contribution to star formation rate  in the inner region is then obtained integrating \ref{sigmastarin} from $R_{acc}$ to $R_{BH}$:

\begin{equation}
\label{sfrinner}
\dfrac{dM_{*,inner}}{dt} = \int_{R_{acc}}^{R_{BH}} 2\pi \dfrac{dM_{BH}}{dt} \dfrac{1}{\pi R_{acc}^2} \left(\dfrac{R}{R_{acc}}\right)^{-\frac{\eta_k}{2(\eta_k-1)}} R dR = \dfrac{dM_{BH}}{dt}\frac{4(\eta_k-1)}{3\eta_k-4} R_{acc}^{\frac{-3\eta_k+4}{2(\eta_k-1)}}\left(R_{BH}^{\frac{3\eta_k-4}{2(\eta_k-1)}}-R_{acc}^{\frac{3\eta_k-4}{2(\eta_k-1)}}\right)
\end{equation}

For the fiducial model, assuming $\eta_k =7/4$ and $R_{acc} \simeq 10^{-2} R_{BH}$, we obtain:

\begin{equation}
\dfrac{dM_{*,inner}}{dt} \simeq 109 \dfrac{dM_{BH}}{dt}~.
\end{equation}
The same calculations can be performed for the outer region using the corresponding $\Sigma_g\propto R^{-\frac{\eta_d+1}{2}}$; here the scaling and the exponent $\alpha_d $ are given in eq. \ref{sigmaoutfiducial}. This yields 

\begin{equation}
\label{sigmastarin3}
\dfrac{d \Sigma_*}{dt} \propto R^{-\frac{\eta_k(\eta_d+1)}{2(\eta_k-1)}}  
\end{equation}

The normalization can be computed imposing continuity between the inner and the outer region; using eq. \ref{sigmastarin} 

\begin{equation}
\label{sigmastarout}
\dfrac{d \Sigma_*}{dt} = \dfrac{dM_{BH}}{dt} \dfrac{1}{\pi R_{acc}^2} \left(\dfrac{R_{BH}}{R_{acc}}\right)^{-\frac{\eta_k}{2(\eta_k-1)}}\left(\dfrac{R}{R_{BH}}\right)^{-\frac{\eta_k(\eta_d+1)}{2(\eta_k-1)}}
\end{equation}

Integrating eq. \ref{sigmastarout} from $R_{BH}$ to $R_0$ we obtain the star formation rate  in the outer region for the fiducial model:

\begin{equation}
\label{sfroutfid}
\dfrac{dM_{*,outer}}{dt} = \dfrac{dM_{BH}}{dt} \dfrac{4(\eta_k-1)}{(-\eta_k \, \eta_d +3 \eta_k -4 )\,R_{acc}^2} \left(\dfrac{R_{BH}}{R_{acc}}\right)^{-\frac{\eta_k}{2(\eta_k-1)}}R_{BH}^{\frac{\eta_k(\eta_d+1)}{2(\eta_k-1)}}\left(R_0^{\frac{- \eta_k \, \eta_d + 3 \eta_k -4}{2 ( \eta_k -1)}}-R_{BH}^{\frac{- \eta_k \, \eta_d + 3 \eta_k -4}{2 ( \eta_k -1)}}\right)
\end{equation}

Assuming $\eta_k=7/4$,  $\eta_d=1/2$, and $R_{acc} \simeq 10^{-2} R_{BH}$ we obtain

\begin{equation}
\dfrac{dM_{*,outer}}{dt} \simeq 289 \dfrac{dM_{BH}}{dt}
\end{equation}

\newpage
\section{BLACK HOLE ACCRETION RATE AND NUCLEAR STAR FORMATION RATE FROM DISK INSTABILITIES: THE CASE OF EXPONENTIAL SURFACE DENSITY PROFILES}

Here we follow the same procedure adopted in Appendix A, changing only the surface density profile (and the the corresponding 
potential) of the galactic disk. In particular, we consider a thin disk whose density profile is described by the following exponential-law:
\begin{equation}
\Sigma_d(R)=\Sigma_0e^{-R/R_d}
\end{equation}
$R_d$ is the exponential length of the disk and its order is of a few kpc. The potential associated to the disk is (Binney$\&$Tremaine, 1987):
\begin{equation}
\Phi(R)=-\pi G \Sigma_0R\left[I_0(y)K_1(y)-I_1(y)K_0(y)\right]
\end{equation}
where $y \equiv R/2R_d$ and $I_n$, $K_n$ are the modified Bessel function of first and second kind. Since both disk and bulge exponential length are much greater than scales of our interest, we can use the asymptotic form for the modified Bessel function; as a result, we obtain
\begin{equation}
\Phi(R) \approx -2 \pi G\Sigma_0R_d
\end{equation}
\begin{equation}
\label{sigma0exp}
\Sigma_d(R) \approx \Sigma_0
\end{equation}
Note that in this case the disk surface density and the potential are computed self-consistently, at variance with the previous case where we assumed a 
WKB form for the potential and a power-law form for the surface density.  

Following the same procedure adopted in Appendix A, we consider two regions: an outer region where the  potential is dominated by the disk, and an inner region where the BH gravity is dominant. The radius $R_{BH}\sim $ 10 pc marks the separations between the two regions, while the scale $R_{acc}\sim 0.1$ pc corresponds to the inner bound of the innermost regions where we want to compute the BH accretion rate. 
The potential of the disk is only needed to calculate the analytic expression for $\Sigma_{g}$ in the outer region, where the BH does not dominate the potential.

In the inner region, the potential is dominated by the BH, and the computation of the BH accretion rate follows exactly the lines 
shown in Appendix A, changing only the scaling of the disk 
radial profile by adopting $\eta_d =0$, as to achieve a constant $\Sigma_d$ as given by eq. \ref{sigma0exp}. 
Thus, the BH accretion rate is given by the same eq. \ref{intern_inflow3} with $\eta_d=0$. However, 
the boundary conditions $A(R_{BH},R_0)$ and $B(R_0)$ since these depend on the gas surface density $\Sigma_{g,outer}$ in the region $R>R_{BH}$ where the potential and the surface densities need to be recomputed (see eq. \ref{boundary1} and \ref{boundary2}). Also, the contribution to the nuclear star formation from 
such a region $R>R_{BH}$ has to be recomputed, since it also depends on $\Sigma_{g,outer}$.

To  compute$\Sigma_{g,outer}$ in the outer region we start from the the perturbed potential in the outer region 
\begin{equation}
|\Phi_a| \approx a_0 f_d \Phi_0 = \dfrac{2a_0 \pi^2 G^2 \Sigma_0^2 R_d }{ \Omega^{2}R}
\end{equation}
with $a_0 \sim 0.3$ and $m=2$, motivated by simulations. Substituting the above expression in eq. \ref{massinflow} we obtain for the outer region (from $R_{BH}$ to $R_0$)
\begin{equation}
\dfrac{\partial}{\partial R} \left(\dfrac{dM_{infl}}{dt}\right) = m a_0 \Sigma_{g}\dfrac{1}{\Omega^{3}R^2} \dfrac{\pi^2 G^2 \Sigma_0^2 R_d }{(2+\nu_{\Omega})}|-\eta_g -3\nu_{\Omega}-1|=
 \alpha_E \pi \Omega \left(\dfrac{\Sigma_0}{\Sigma_{t}}\right)^2 \Sigma_{g} R_d 
\end{equation}
with
\begin{equation}
\nu_u \equiv  \dfrac{\partial \, ln \, u}{\partial \, ln \, R}
~~~~~~~~~~~~~~~{\rm and}~~~~~~~~~~~~~~
\alpha_E \equiv \dfrac{ma_0}{\pi} \dfrac{|-\eta_g -3\nu_{\Omega}-1|}{(2+\nu_{\Omega})}
\end{equation}
where we have used the relation $\Omega^2\,R=G\,\int 2\,\pi\,\Sigma_t\,R\,dR$, and $\Sigma_t$ is the total (disk + bulge) surface density profiles 
(see Appendix A). Substituting such an expression in eq. \ref{masstot} yields 
\begin{equation}
\label{sigmagoutexp}
\left(\dfrac{\Sigma_{g}}{\Sigma_{k}}\right)_{outer}^{\eta_k-1} = \dfrac{\alpha_E }{2} \left(\dfrac{\Sigma_0}{\Sigma_{t}}\right)^2 \dfrac{\Omega t_k R_d}{R} 
\end{equation}

That is the equation for $\Sigma_{g}$ in the outer region $R_{BH}\leq R\leq R_0$, the counterpart of eq. \ref{sigmaoutfiducial} in the case of an 
exponential disk surface density profile;  in the inner region $R_{acc}\leq R\leq R_{BH}$, the analytic expression for $\Sigma_{g}$ is the same of eq. \ref{sigmain}, since this region is dominated by BH potential. 
We can now recompute the boundary conditions $A(R_{BH},R_0)$ and $B(R_0)$. To this aim, 
we use eqs. \ref{boundary1} and \ref{boundary2}, but adopting for the gas surface density in the outer region $\Sigma_{g,outer}$ 
the expression in eq. \ref{sigmagoutexp} valid for the case of an exponential disk. This yields

\begin{align}
A(R_{BH},R_0)=  \left(1+ \dfrac{4 \eta_k-7- \eta_d}{4 \eta_k - 5} \left(\dfrac{\Sigma_0}{\Sigma_t}\dfrac{R_d}{R_{BH}} \dfrac{\alpha_E}{2 \alpha_{BH}}\right)^{-\frac{1}{\eta_k-1}}/B(R_0)\right)^{-1} & \hspace{0cm}
\end{align}
In practice, the right-hand term in the parenthesis is extremely small, due to the choice of $\eta_k$, $\eta_d$ ( $\eta_d \sim 0$ since $\Sigma_0$ is almost constant) and to the factor $R_d/R_{BH}$, so that the above expression for the case of an exponential disk yields always $A(R_{BH},R_0)\approx 1$, independently on the outer boundary condition, which anyway reads  (in the case $\eta_k=7/4$)
\begin{align}
\label{div3}
B(R_0) \approx \Bigg[   1+26\,{f_0\over f_{gas}}   \Bigg({R_d\over R_0 }  \Bigg)^{4/3}     \Bigg]^{-1}
\end{align}
where $f_0$ and $f_{gas}$ are those given in eq. \ref{boundary2}. Thus, in the main text, we shall adopt the excellent approximation $A(R_{BH},R_0)=1$.

As for the nuclear star formation, the contribution from the inner region $R_{acc}\leq R\leq R_{BH}$ is the same given in eq. \ref{sfrinner}, since 
in this region the potential is dominated by the BH. In the outer region, we follow the procedure presented in sect. A5, based on the relation 
$\Sigma_{*,outer}\propto \Sigma_{g,outer}^{\eta_k}$, with the latter quantity taken from eq. \ref{sigmagoutexp}. This yields  

\begin{equation}
\label{sigmastarout2}
\dfrac{d \Sigma_*}{dt} = \dfrac{dM_{BH}}{dt} \dfrac{1}{\pi R_{acc}^2} \left(\dfrac{R_{BH}}{R_{acc}}\right)^{-\frac{\eta_k}{2(\eta_k-1)}}\left(\dfrac{R}{R_{BH}}\right)^{-\frac{\eta_k(\eta_d+3)}{2(\eta_k-1)}}
\end{equation}
Integrating eq. \ref{sigmastarout2} from $R_{BH}$ to $R_0$ we obtain 

\begin{equation}
\dfrac{dM_{*,outer}}{dt} = \dfrac{dM_{BH}}{dt} \dfrac{4(\eta_k-1)}{(-\eta_k \, \eta_d + \eta_k -4 )\,R_{acc}^2} \left(\dfrac{R_{BH}}{R_{acc}}\right)^{-\frac{\eta_k}{2(\eta_k-1)}}R_{BH}^{\frac{\eta_k(\eta_d+3)}{2(\eta_k-1)}}\left(R_0^{\frac{- \eta_k \, \eta_d +  \eta_k -4}{2 ( \eta_k -1)}}-R_{BH}^{\frac{- \eta_k \, \eta_d +  \eta_k -4}{2 ( \eta_k -1)}}\right)
\end{equation}

Taking $\eta_k=7/4$, $\eta_d=0$ and $R_0=100$ pc

\begin{equation}
\dfrac{dM_{*,outer,exp}}{dt} = \dfrac{4}{3}\dfrac{dM_{BH}}{dt} R_{BH}^{\frac{7}{3}}R_{acc}^{-\frac{5}{6}}\left(R_{BH}^{-\frac{3}{2}}-R_{0}^{-\frac{3}{2}}\right) \simeq 60 \dfrac{dM_{BH}}{dt}
\end{equation}
Note that, at variance with the case of power-law profile given in eq. \ref{sfroutfid}, the star formation rate does not diverge with increasing 
$R_0$, due to the exponential decay of the disk surface density.


\newpage
\begin{thebibliography}{}
\bibitem{}Abazajian, K. N., Adelman-McCarthy, J. K., Agüeros, M. A., et al. 2009, ApJS, 182, 543
\bibitem{}Adelman-McCarthy J. K. et al., 2006, ApJS, 162, 38
\bibitem{}Aird, J., Nandra, K., Laird, E.S., et al. 2010, MNRAS, 401, 2531
\bibitem{}Aller M. C., Richstone D. O., 2007, ApJ, 665, 120
\bibitem{}Alonso M. S., Lambas D. G., Tissera P. B., Coldwell G., 2007, MNRAS, 375, 1017
\bibitem{}Angl\'es-Alc\'azar, D., \"Ozel, F., Dav\'e, R. 2013, ApJ, 770, 5
\bibitem{}Anderson, C.S., Johnston, H.M., Hunstead, R.W. 2013, MNRAS, 431, 3269
\bibitem{}Athanassoula, E. 1992, MNRAS, 259, 328
\bibitem{}Bahcall, J. N., Kirhakos, S., Saxe, D. H., Schneider, D. P. 1997, ApJ, 479, 642
\bibitem{}Baldry, I.K., Glazebrook, K., Brinkmann, J., Zeljko, I., Lupton, R.H., Nichol, R.C., Szalay, A.S. 2004, ApJ, 600, 681 
\bibitem{}Baldry, I.K. et al., 2012, MNRAS, 421, 621
\bibitem{}Bardeen J. M., Bond J. R., Kaiser N., Szalay A. S., 1986, ApJ, 304, 15 
\bibitem{}Barnes, J.E., Hernquist, L.E. 1991, ApJ, 370, L65
\bibitem{}Baugh, C.M. 2006, Rep. Prog. Phys. 69, 3101
\bibitem{}Bennert, N., Canalizo, G., Jungwiert, B., Stockton, A., Schweizer, F., Peng, C. Y., Lacy, M. 2008, ApJ, 677, 846
\bibitem{}Benson A. J., Devereux N., 2010, MNRAS, 402, 2321
\bibitem{}Binney J., Tremaine S., 1987, Galactic dynamics. Princeton Univ. Press, Princeton, p. 747
\bibitem{}Bond, J.R., Cole, S., Efstathiou, G., \& Kaiser, N.,1991, ApJ, 379, 440
\bibitem{}Bongiorno, A., Zamorani, G., Gavignaud, I., et al. 2007, A\&A, 472, 443
\bibitem{}Bouche', N. et al. 2007, ApJ, 671, 303
\bibitem{}Bournaud, F., Combes, F.,  Semelin, B. 2005, MNRAS, 364, L18
\bibitem{}Bournaud, F., Elmegreen, B. G.,  Elmegreen, D. M. 2007, ApJ, 670, 237
\bibitem{}Bournaud, F. et al., 2011, ApJ, 741, L33
\bibitem{}Bournaud, F. et al., 2012, ApJ, 757, 81
\bibitem{}Bouwens, R.G., Illingworth, G.D., Franx, M., Ford, H. 2007, ApJ, 670, 928
\bibitem{}Bouwens, R.J. et al. 2011, ApJ, 737, 90
\bibitem{}Bower, R.G. 1991, MNRAS, 248, 332
\bibitem{}Bower, R.G., Benson, A.J., Malbon, R., Helly, J.C., Frenk, C.S., Baugh, C.M., Cole, S., Lacey, C.G. 2006, MNRAS, 370, 645
\bibitem{}Brusa, M., Civano, F., Comastri, A., et al. 2010, ApJ, 716, 348
\bibitem{} Bruzual, G., Charlot, S. 2003, MNRAS, 344, 1000
\bibitem{}Cappellari, M. et al. 2006, MNRAS, 366, 1126
\bibitem{}Cavaliere, A.,  \& Vittorini, V.  2000, ApJ, 543, 599
\bibitem{}Ceverino, D., Bournaud, F., Dekel, A. 2010, MNRAS, 404, 2151
\bibitem{}Cirasuolo M., McLure R. J., Dunlop J. S., Almaini O., Foucaud S., Simpson C., 2010, MNRAS, 401, 1166
\bibitem{}Cisternas M. et al. 2011, ApJ, 726, 57
\bibitem{}Civano, F., Brusa, M., Comastri, A. et al. 2011 ApJ, 741, 91
\bibitem{}Cole S., Lacey C. G., Baugh C. M., Frenk C. S., 2000, MNRAS, 319, 168
\bibitem{}Combes, F., \& Gerin, M. 1985, A\&A, 150, 327
\bibitem{}Combes, F. et al. 2009, A\&A, 503, 73
\bibitem{}Croton D. J. et al., 2006, MNRAS, 365, 11
\bibitem{}Cox, T. J., Dutta, S. N., Di Matteo, T., Hernquist, L., Hopkins, P. F., Robertson, B.,  Springel, V. 2006, ApJ, 650, 791
\bibitem{}Cox, T.J. et al. 2008, MNRAS, 384, 386
\bibitem{}Croom, S.M., Richards, G.T., Shanks, T., et al. 2009, MNRAS, 399, 1755
\bibitem{}Daddi, E., Elbaz, D., Walter, F., et al. 2010, ApJ, 714, L118
\bibitem{}Darg D. W. et al., 2010, MNRAS, 401, 1552
\bibitem{}Davies, R.I.,Sa ́nchez, F.M.,Genzel, R.,Tacconi, L.J., Hicks, E.K.S., Friedrich, S., Sternberg, A., 2007, ApJ, 671, 1388
\bibitem{}De Lucia G., Fontanot F., Wilman D., Monaco P., 2011, MNRAS, 414, 1439
\bibitem{}de Vries, W. H., Morganti, R., Ro ̈ttgering, H. J. A., Vermeulen, R., van Breugel,
W., Rengelink, R., \& Jarvis, M. J. 2002, AJ, 123, 1784
\bibitem{}Disney, M.J., et al. 1995, Nature, 376, 150
\bibitem{}Dotti, M., Ruszkowski, M., Paredi, L., Colpi, M., Volonteri, M.,
 Haardt, F. 2009, MNRAS, 396, 1640
\bibitem{}Drory, N. et al. 2004, ApJ, 608, 742
\bibitem{}Ebrero, J., Carrera, F. J., Page, M. J., et al. 2009, A\&A, 493, 55
\bibitem{}Efstathiou G., Lake G., Negroponte J., 1982, MNRAS, 199, 1069
\bibitem{}Eisenhardt, P. R., et al. 2004, ApJS, 154, 48
\bibitem{}Eliche-Moral, M. C., Gonza ́lez-Garc ́ıa, A. C., Balcells, M., Aguerri, J. A. L., Gallego, J.,  Zamorano, J. 2008, in ASP Conf. Ser. 396, Formation and Evolution of Galaxy Disks, ed. J. G. Funes  E. M. Corsini (San Francisco, CA: ASP), 359
\bibitem{}Ellison, S.L., Patton, D.R., Mendel, J.T., Scudder, J.M. 2011, MNRAS, 418, 2043
\bibitem{}Fanidakis, N. et al. 2012, MNRAS, 419, 2797
\bibitem{}Ferrarese, L. Merritt, D., 2000, ApJ, 539, L9
\bibitem{}Fiore, F., et al. 2012, A\&A, 537, 22
\bibitem{}Fisher, D.B., Drory, N. 2011, ApJ, 733, L47
\bibitem{}Fontana A. et al., 2006, A\&A, 459, 745
\bibitem{}Fontanot, F., Cristiani, S., Monaco, P., Nonino, M., Vanzella, E., Brandt, W. N., Grazian, A., \& Mao, J. 2007, A\&A, 461, 39
\bibitem{}Friedli, D., \& Martinet, L. 1993, A\&A, 277, 27
\bibitem{}Gebhardt, K. et al. 2000, ApJ, 539, L13
\bibitem{}Genzel, R., et al. 2008, ApJ, 687, 59
\bibitem{}Genzel, R., Tacconi, L. J., Gracia-Carpio, J., et al. 2010, MNRAS, 407, 2091
\bibitem{}Giavalisco, M. et al. 2004, ApJ, 600, L103
\bibitem{}Glikman, E., Djorgovski, S.G., Stern, D., Dey, A., Jannuzi, B.T., \& Lee, K.-S. 2011, \apj, 728, L26
\bibitem{}Giallongo, E., Menci, N., Fiore, F., Castellano, M., Fontana, A., Grazian, A., Pentericci, L. 2012, ApJ, 755, 124
\bibitem{}Goldreich, P.,  Lynden-Bell, D. 1965, MNRAS, 130, 97
\bibitem{}Goldreich, P.,  Tremaine, S. 1980, ApJ, 241, 425
\bibitem{}Gonzalez Delgado, R.M. Arribas, S., Perez E., Heckman, T. 2002, ApJ, 579 188
\bibitem{}Grogin, N.A. et al. 2005, ApJ 627, L97
\bibitem{}H\"aring, N., Rix, H.-W. 2004, ApJ, 604, L89
\bibitem{}Heller, C.H., \& Shlosman, I. 1994, ApJ, 424, 84
\bibitem{}Hennawi, J. F., et al. 2010, ApJ, 719, 1672
\bibitem{}Henriques, B.M.B., \& Thomas, P.A. 2010, MNRAS, 403, 768  
\bibitem{}Hernquist, L. 1989, Nature, 340, 687
\bibitem{}Hernquist, L.,  Mihos, J.C. 1995, ApJ, 448, 41
\bibitem{}Hickox, R.C. et al. 2009, ApJ, 696, 891
\bibitem{}Hicks E. K. S., Davies R. I., Malkan M. A., Genzel R., Tacconi L. J., Sa ́nchez
F. M., Sternberg A., 2009, ApJ, 696, 448
\bibitem{}Hildebrandt, H., Pielorz, J., Erben, T., et al. 2009, A\&A, 498, 725
\bibitem{}Hirschmann, M., Somerville, R.S., Naab, T., Burkert, A. 2012, MNRAS, 426, 237
\bibitem{}Hopkins, P. F., Richards, G. T.,  Hernquist, L. 2007, ApJ, 654, 731
\bibitem{}Hopkins, P.H. et al. 2009a, MNRAS, 397, 802
\bibitem{}Hopkins, P.F., Quataert, E. 2010, MNRAS, 407, 1529
\bibitem{}Hopkins, P.F., Quataert, E. 2011, MNRAS, 411, 1027 (HQ11)
\bibitem{}Hutchings, J.B. 1987, ApJ, 320, 122
\bibitem{}Jiang, L., Fan, X., Bian, F. et al. 2009, AJ, 138, 305
\bibitem{}Jogee, S. 2006, Lect. Notes Phys, 693, 143
\bibitem{}Jones, D.H., Peterson, B.A., Colless, M., Saunders, W. 2006, MNRAS, 396, 535
\bibitem{}Kaspi, S., Smith, P. S., Netzer, H., Maoz, D., Jannuzi, B. T., Giveon, U.
2000, ApJ, 533, 631
\bibitem{}Kauffmann G. et al., 2003b, MNRAS, 346, 1055
\bibitem{}Kauffmann G., Heckman T. M., 2009, MNRAS, 397, 135 
\bibitem{}Kenter, A., et al. 2005, ApJS, 161, 9
\bibitem{}Khochfar S., Silk J., 2006, MNRAS, 370, 902
\bibitem{}King I., 1962, AJ, 67, 471
\bibitem{}Kirhakos, S., Bahcall, J.N., Schneider, D.P., Kristian, J. 1999, ApJ, 520, 67
\bibitem{}Kocevski D. D., Faber S. M., Mozena M., Koekemoer
A. M., Nandra K., Rangel C., Laird E. S., Brusa M., Wuyts S., Trump J. R., Koo D. C., Somerville R. S., Bell E. F., Lotz J. M., Alexander D. M., Bournaud F., Conselice C. J., Dahlen T., 2012, ApJ, 744, 148
\bibitem{}Kollatschny, W., Reichstein, A., Zetzl, M. 2012, A\&A, 548, 37
\bibitem{}Kollmeier J. A. et al., 2006, ApJ, 648, 128
\bibitem{}Kormendy, J., Richstone, D. 1995, ARA\&A, 33, 581
\bibitem{}Koss M., Mushotzky R., Veilleux S., Winter L., 2010, ApJ, 716, L125
\bibitem{}Lacey, C., \& Cole, S., 1993, MNRAS, 262, 627 
\bibitem{}La Franca, F. et al. 2005, ApJ, 635, 864
\bibitem{}Lamastra, A., Menci, N., Maiolino, R. Fiore, F., Merloni, A. 2010, MNRAS, 405, 29
\bibitem{}Lamastra, A., Menci, N., Fiore, F., Santini, P. 2013a, A\&A, 552, 44
\bibitem{}Lamastra, A., Menci, N., Fiore, F., et al. 2013b, A\&A, 559, A56
\bibitem{}Li C., White S. D. M., 2009, MNRAS, 398, 2177
\bibitem{}Li C., Kauffmann G., Wang L., White S. D. M., Heckman T. M., Jing Y. P., 2006, MNRAS, 373, 457
\bibitem{}Lilly, S. J., Le Fèvre, O., Renzini, A., et al. 2007, ApJS, 172, 70
\bibitem{}Lin, C. C., Yuan, C.,  Shu, F. H. 1969, ApJ, 155, 721
\bibitem{}Lotz J. M., Jonsson P., Cox T. J., Croton D., Primack J. R., Somerville R.
\bibitem{}Lynden-Bell, D.,  Kalnajs, A. J. 1972, MNRAS, 157, 1
\bibitem{}Maciejewski, W., Sparke, L.S. 2000, MNRAS, 313, 745
\bibitem{}Mayer, L., Kazantzidis, S., Madau, P., Colpi, M., Quinn, T., Wadsley, J. 2007, Science, 316, 1874
\bibitem{}Magorrian, J. et al. 1998, AJ, 115, 2285
\bibitem{}Makino J., Hut P., 1997, ApJ, 481, 83
\bibitem{}Marconi, A., Hunt, L.K., 2003, ApJ, 589, L21
\bibitem{}Marconi, A., Risaliti, G., Gilli, R., Hunt, L.K.,Maiolino, R., Salvati, M. 2004, MNRAS, 351, 169
\bibitem{}McGurk R. C., Max C. E., Rosario D. J., Shields G. A., Smith K. L., Wright, S. A., 2011, ApJ, 738, 2
\bibitem{}Menci, N., Cavaliere, A., Fontana, A., Giallongo, E., Poli, F., Vittorini, V. 2003, ApJ, 587, L63
\bibitem{}Menci, N., Cavaliere, A., Fontana, A., Giallongo, E., \& Poli, F. 2004,ApJ, 604, 12
\bibitem{}Menci, N., Fontana, A., Giallongo, E., Grazian, A.,Salimbeni, S. 2006, ApJ,  647, 753
\bibitem{}Menci, N., Fiore, F., Puccetti, S., Cavaliere, A. 2008, ApJ, 686, 219
\bibitem{}Mo, H.J, Mao S., \& White, S.D.M., 1998, MNRAS, 295, 319 
\bibitem{}Murray, S. S., et al. 2005, ApJS, 161, 1
\bibitem{}Navarro, J. F., Frenk, C. S., White, S. D. M. 1997, ApJ, 490, 493
\bibitem{}Netzer, H. 2009a, MNRAS, 399, 1907
\bibitem{}Netzer, H. 2009b, ApJ, 695, 793
\bibitem{}Ostriker, E. C., Shu, F. H., Adams, F. C. 1992, ApJ, 399, 192
\bibitem{}Panter, B., Jimenez, R., Heavens, A.F., Charlot, S. 2007, MNRAS, 378, 1150
\bibitem{}Parry O. H., Eke V. R., Frenk C. S., 2009, MNRAS, 396, 1972
\bibitem{}Pierce C. M.et al. 2007, ApJ, 660, L19
\bibitem{}Reddy, N.A., Steidel, C.S. 2009, ApJ,  692, 778
\bibitem{}Richards, G.T. et al. 2006, ApJ, 131, 2766
\bibitem{}Robertson, B., Cox, T. J., Hernquist, L., Franx, M., Hopkins, P. F.,
Martini, P., Springel, V. 2006a, ApJ, 641, 21
\bibitem{}Robertson, B., Hernquist, L., Cox, T. J., Di Matteo, T., Hopkins, P. F., Martini, P., Springel, V. 2006b, ApJ, 641, 90
\bibitem{}Rodighiero, G., Daddi, E., Baronchelli, I., et al. 2011, ApJ, 739, L40
\bibitem{}Rogers B., Ferreras I., Kaviraj S., Pasquali A., Sarzi M., 2009, MNRAS, 399, 2172
\bibitem{}Rosario, D.J. et al. 2013a, 763, 59
\bibitem{}Rosario, D.J. et al 2013b, A\&A, 560, A72
\bibitem{}Salucci P., Szuszkiewicz E., Monaco P., Danese L., 1999, MNRAS, 307, 637
\bibitem{}Sanders, D. B., \& Mirabel, I.F. 1996, ARA\&A, 34, 749
\bibitem{}Santini, P. et al. 2014, A\&A, 562, 30
\bibitem{}Saslaw, W.C., 1985, {\it  Gravitational Physics of Stellar and Galactic Systems} (Cambridge: Cambridge Univ. Press) 
\bibitem{}Schwarz, M.P. 1984, MNRAS, 209, 93
\bibitem{}Shankar F., Weinberg D. H., \& Miralda-Escud´e J. 2009, ApJ, 690, 20
\bibitem{}Shankar F., Marulli, F., mathur, S. Bernardi, M., Bournaud, F. 2012, A\&A, 540, A23
\bibitem{}Shankar F., Weinberg D. H., \& Miralda-Escud´e J. 2013, MNRAS, 428, 421
\bibitem{}Shirasaki, Y., Tanaka, M., Ohishi, M., et al. 2011, PASJ, 63, 469
\bibitem{}Shlosman, I., Frank, J., \& Begelman, M.C. 1989, Nature, 338, 45
\bibitem{}Shu, F. H., Tremaine, S., Adams, F. C., Ruden, S. P. 1990, ApJ, 358, 495
\bibitem{}Siana, B., Polletta, M., Smith, H.E. et al. 2008, \apj, 675, 49
\bibitem{}Silverman, J.D. et al., 2009, ApJ, 696,396
\bibitem{}Silverman, J.D. et al. 2011, ApJ, 743, 2
\bibitem{}Somerville R. S., Primack J. R., 1999, MNRAS, 310, 1087
\bibitem{}Somerville R. S., Gilmore, R.C., Primack J. R., Dominguez, A. 2012, MNRAS, 423, 1992
\bibitem{}Spergel, D.N. et al. 2006, ApJ,  in  press (astro-ph/0603449)
\bibitem{}Stewart K. R., 2009, in Jogee S., Marinova I., Hao L., Blanc G. A., eds, ASP
Conf. Ser. Vol. 419, Galaxy Evolution: Emerging Insights and Future
Challenges. Astron. Soc. Pac. San Francisco, p. 243
\bibitem{}Storchi-Bergmann T., Gonzalez-Delgado R. M., Schmitt H. R., Cid-
\bibitem{}Taylor J. E., Babul A., 2001, ApJ, 559, 716
\bibitem{}Taniguchi, Y. 1999, ApJ, 524, 65
\bibitem{}Teyssier R., Chapon D., Bournaud F., 2010, ApJ, 720, L149
\bibitem{}Toomre, A. 1969, ApJ, 158, 899
\bibitem{}Trakhtenbrot, B., Netzer, H. 2012, MNRAS, 427, 3081
\bibitem{}Treister, E., Schawinski, K., Urry, C.M., Simmons, B.D. 2012, ApJ, 758, L39
\bibitem{}Tremaine S. et al., 2002, ApJ, 574, 740
\bibitem{}Trump, J. R., Impey, C. D., Elvis, M., et al. 2009, ApJ, 696, 1195
\bibitem{}Ueda, Y. et al. 2014, ApJ, in press
\bibitem{}van der Burg, R.F.J., Hildebrandt, H.,  Erben, T. 2010, A\&A, 523, 74
\bibitem{}Villforth, C. et al. 2014, MNRAS, 439, 3342
\bibitem{}Villar-Martín, M. et al. 2010, MNRAS, 416, 262
\bibitem{}Villar-Martín, M. et al. 2012, MNRAS, 423, 80
\bibitem{}Woods D. F., Geller M. J., 2007, AJ, 134, 527
\bibitem{}Yates, M.G., Miller, L., \& Peacock, J.A. 1989, MNRAS, 240, 129
\bibitem{}York D. G. et al., 2000, AJ, 120, 1579
\bibitem{}Younger, J. D., Hopkins, P. F., Cox, T. J.,  Hernquist, L. 2008, ApJ, 686, 815
\bibitem{}Yu, Q., \& Tremaine, S., 2002, MNRAS, 335, 965













\end{thebibliography}
\end{document}